\documentclass[sigconf,balance=false,authorversion]{acmart} %

\usepackage{popets}
\setcopyright{popets}
\copyrightyear{2024}
\acmYear{2024}
\acmVolume{2024}
\acmNumber{3}
\acmDOI{10.56553/popets-2024-0068}
\acmConference{Privacy Enhancing Technologies}
\makeatletter
\newif\ifreview
\@ifclasswith{acmart}{review}{\reviewtrue}{\reviewfalse}
\makeatother

\usepackage{algorithmic}
\usepackage{textcomp}
\usepackage{tabularx}
\newcolumntype{Y}{>{\centering\arraybackslash}X} %
\usepackage{xcolor}
\usepackage[inline]{enumitem}
\usepackage{dsfont}
\usepackage{xspace}
\usepackage{multirow}
\usepackage{pifont}
\newcommand{\xmark}{\ding{55}\xspace}
\usepackage{tikz}
\usepackage[binary-units]{siunitx} %
\newlist{questions}{enumerate}{1}
\setlist[questions,1]{label=\textbf{RQ\arabic*.},ref=\textbf{RQ\arabic*}}
\usepackage[section,numberedsection=autolabel,nogroupskip,nonumberlist]{glossaries}
\makeglossaries  %
\newacronym{conv1d}{Conv1D}{1D Convolution}
\newacronym{aae}{AAE}{Adversarial Autoencoder}
\newacronym{ae}{AE}{Autoencoder}
\newacronym{bce}{BCE}{Binary Cross Entropy}
\newacronym{ce}{CE}{Cross Entropy}
\newacronym{cnn}{CNN}{Convolutional Neural Network}
\newacronym{dl}{DL}{Deep Learning}
\newacronym{dp-stg}{DP-STG}{Differentially Private Synthetic Trajectory Generator}
\newacronym{dp}{DP}{Differential Privacy}
\newacronym{dtw}{DTW}{Dynamic Time Warping}
\newacronym{em}{EM}{Exponential Mechanism}
\newacronym{emd}{EMD}{Earth Mover's Distance}
\newacronym{exGAN}{exGAN}{Except-Condition GAN}
\newacronym{fs}{FS-NYC}{Foursquare NYC~\cite{fs_nyc}}
\newacronym{gan}{GAN}{Generative Adversarial Network}
\newacronym{geo-ind}{Geo-Ind}{Geo-Indistinguishability}
\newacronym{gru}{GRU}{Gated Recurrent Unit}
\newacronym{hd}{HD}{Hausdorff Distance}
\newacronym{jsd}{JSD}{Jensen Shannon Distance}
\newacronym{lldp}{LLDP}{Local Label Differential Privacy}
\newacronym{ldp}{LDP}{Local Differential Privacy}
\newacronym{lstm}{LSTM}{Long Short-Term Memory}
\newacronym{mae}{MAE}{Mean Absolute Error}
\newacronym{mi}{MI}{Mutual Information}
\newacronym{mia}{MIA}{Membership Interference Attack}
\newacronym{mlp}{MLP}{Multi-Layer Perceptron}
\newacronym{mse}{MSE}{Mean Squared Error}
\newacronym{mwe}{MWE}{Minimal Working Example}
\newacronym{nlp}{NLP}{Natural Language Processing}
\newacronym{poi}{POI}{Point of Interest}
\newacronym{RAoPT}{RAoPT}{Reconstruction Attack on Protected Trajectories}
\newacronym{rdp}{RDP}{Rényi Differential Privacy}
\newacronym{relu}{ReLU}{Rectified Linear Activation}
\newacronym{rnn}{RNN}{Recurrent Neural Network}
\newacronym{sdd}{SDD}{Sampling Distance and Direction}
\newacronym{stc}{STC}{Spatial-Temporal-Categorical Distance}
\newacronym{stg}{STG}{Synthetic Trajector Generator}
\newacronym{stn}{STN}{Large Spatial Transformer Network}
\newacronym{tanh}{tanh}{Hyperbolic Tangent}
\newacronym{TRA}{TRA}{Trajectory Reconstruction Attack}
\newacronym{TSG}{TSG}{Two-Stage-\gls{gan}}
\newacronym{tul}{TUL}{Trajectory User Linking}
\newacronym{uop}{UoP}{Unit of Privacy}
\newacronym{vae}{VAE}{Variational Autoencoder}
\newacronym{wd}{WD}{Wasserstein Distance}
\newacronym{wgan}{WGAN}{Wasserstein \gls{gan}}
\newacronym{tcac}{TCAC}{Trajectory Category Auxiliary Classifier}
\newglossaryentry{fc}{
    name={FC},
    short={FC},
    description={Fully Connected layer. Also called \textit{Dense} or \textit{Linear} layer},
    first={Fully Connected (FC)},
    long={Fully Connected}
}
\newglossaryentry{dp-sgd}{
    name={DP-SGD},
    short={DP-SGD},
    first={Differentially Private Stochastic Gradient Descent (DP-SGD)},
    long={Differentially Private Stochastic Gradient Descent},
    description={Differentially Private Stochastic Gradient Descent, refer \secref{sec_dp-sgd}}
}
\newglossaryentry{wgan-lp}{
    name={WGAN-LP},
    short={WGAN-LP},
    first={\glsentryshort{wgan} with Lipschitz Penalty (WGAN-LP)},
    description={\glsentryshort{wgan} with Lipschitz Penalty \cite{Petzka2018}}
}
\newglossaryentry{wgan-gp}{
    name={WGAN-GP},
    short={WGAN-GP},
    first={\glsentryshort{wgan} with Gradient Penalty (WGAN-GP)},
    description={\glsentryshort{wgan} with Gradient Penalty \cite{iWGAN}}
}
\newglossaryentry{mnist-seq}{
    name={MNIST-Seq},
    short={MNIST-Seq},
    first={MNIST Sequential (MNIST-Seq)},
    description={MNIST Sequential Dataset: Images are transformed to sequences of length $28$ with $28$ features each \cite{Esteban2017}}
}
\newglossaryentry{gpg}{
    name={GPG},
    short={GPG},
    first={GeoPointGAN (GPG)},
    description={GeoPointGAN~\cite{GeoPointGAN}}
}
\newglossaryentry{adamw}{
    name={AdamW},
    short={AdamW},
    long={AdamW},
    first={AdamW},
    description={PyTorch's AdamW Optimizer}
}
\newacronym{gtg}{GTG}{GeoTrajGAN}
\newacronym{ntg}{NTG}{Noise-TrajGAN}
\newacronym{ar}{AR-RNN}{Autoregressive \glsentryshort{rnn}}
\newacronym{start}{START-RNN}{Start-Point \glsentryshort{rnn}}
\newacronym{rgan}{RGAN}{Recurrent \glsentryshort{gan}}

\makeatletter
\newif\ifanonymous
\@ifclasswith{acmart}{anonymous}{\anonymoustrue}{\anonymousfalse}
\makeatother

\DeclareMathOperator*{\argmin}{argmin}
\makeatletter\let\myfsize\f@size\makeatother
\newcommand{\subheading}[1]{
    \noindent{\textbf{#1.}}
    \addcontentsline{toc}{subsubsection}{#1}
}

\newcommand{\probP}{\text{I\kern-0.15em P}}  %

\newcommand{\refer}{ref.\  }
\newcommand{\secref}[1]{\hyperref[#1]{Section~\ref*{#1}}}
\newcommand{\figref}[1]{\hyperref[#1]{Figure~\ref*{#1}}}
\newcommand{\tabref}[1]{\hyperref[#1]{Table~\ref*{#1}}}

\newcounter{goal}
\renewcommand*\thegoal{G\arabic{goal}}
\newcommand{\gref}[1]{\hyperref[#1]{\textbf{\ref*{#1}: \csname goalname#1\endcsname}}\xspace}
\newcommand{\grefshort}[1]{\textbf{\ref{#1}}\xspace}
\newcommand{\goal}[2]{%
    \refstepcounter{goal}%
    \subsection*{\thegoal: #1}\label{#2}%
    \addcontentsline{toc}{subsection}{\thegoal: #1}%
    \expandafter\gdef\csname goalname#2\endcsname{#1}%
}

\newcommand{\pitfall}[1]{\hyperref[pitfall:#1]{\textbf{Pitfall~\ref*{pitfall:#1}}\xspace}}

\newcommand{\ballnumber}[1]{%
    \tikz[baseline={(myanchor.base)}]{
        \node[circle, fill=., inner sep=0pt, minimum width=1em, minimum height=1em, text centered] (myanchor) {\color{-.}\bfseries\footnotesize #1};
    }
}

\begin{document}

\title[]{SoK: Can Trajectory Generation Combine Privacy and Utility?}

\author{Erik Buchholz}
\orcid{0000-0001-9962-5665}
\affiliation{%
  \institution{University of New South Wales}
  \institution{CSIRO's Data61, Cyber Security CRC}
  \city{Sydney}
  \state{NSW}
  \country{Australia}
}
\email{e.buchholz@unsw.edu.au}

\author{Alsharif Abuadbba}
\orcid{0000-0001-9695-7947}
\affiliation{%
  \institution{CSIRO's Data61}
  \institution{Cyber Security CRC}
  \city{Sydney}
  \state{NSW}
  \country{Australia}
}
\email{sharif.abuadbba@data61.csiro.au}

\author{Shuo Wang}
\orcid{0000-0001-8938-2364}
\affiliation{%
  \institution{CSIRO's Data61}
  \institution{Cyber Security CRC}
  \city{Sydney}
  \state{NSW}
  \country{Australia}
}
\email{shuo.wang@data61.csiro.au}

\author{Surya Nepal}
\orcid{0000-0002-3289-6599}
\affiliation{%
 \institution{CSIRO's Data61}
 \institution{Cyber Security CRC}
 \city{Sydney}
 \state{NSW}
 \country{Australia}
 }
 \email{surya.nepal@data61.csiro.au}

\author{Salil S. Kanhere}
\orcid{0000-0002-1835-3475}
\affiliation{%
  \institution{University of New South Wales}
  \city{Sydney}
  \state{NSW}
  \country{Australia}
}
\email{salil.kanhere@unsw.edu.au}

\begin{abstract}
While location trajectories represent a valuable data source for analyses and location-based services, they can reveal sensitive information, such as political and religious preferences.
Differentially private publication mechanisms have been proposed to allow for analyses under rigorous privacy guarantees.
However, the traditional protection schemes suffer from a limiting privacy-utility trade-off and are vulnerable to correlation and reconstruction attacks.
Synthetic trajectory data generation and release represent a promising alternative to protection algorithms.
While initial proposals achieve remarkable utility, they fail to provide rigorous privacy guarantees.
This paper proposes a framework for designing a privacy-preserving trajectory publication approach by defining five design goals, particularly stressing the importance of choosing an appropriate Unit of Privacy.
Based on this framework, we briefly discuss the existing trajectory protection approaches, emphasising their shortcomings.
This work focuses on the systematisation of the state-of-the-art generative models for trajectories in the context of the proposed framework.
We find that no existing solution satisfies all requirements.
Thus, we perform an experimental study evaluating the applicability of six sequential generative models to the trajectory domain.
Finally, we conclude that a generative trajectory model providing semantic guarantees remains an open research question and propose concrete next steps for future research.
\end{abstract}

\keywords{Trajectory Privacy, Differential Privacy, Location Privacy, Deep Learning, Generative Adversarial Networks}

\maketitle

\section{Introduction}

Location trajectories are valuable for several applications, from navigation services over research to pandemic control.
Moreover, a large amount of location data is collected daily through the increasing number of sensor-equipped devices, particularly smartphones.
However, the information content of location trajectories poses a risk of exposing sensitive details about individuals, such as religious, political, or sexual beliefs~\cite{Primault2019, Abul2008}.
For instance, researchers showed in 2013 that only four locations provided through the connection to mobile phone antennas suffice to uniquely identify \SI{95}{\%} of users~\cite{DeMontjoye2013}.
In the case of an anonymised dataset released by a New York taxi company, a Redditor identified which taxi drivers are practising Muslims by correlating the drivers' break times with mandatory prayer times~\cite{Franceschi-Bicchierai}.
Such examples illustrate that mobility datasets require appropriate protection when released. 

Various protection mechanisms based on $k$-Anonymity~\cite{Sweeney2002} and \gls{dp}~\cite{Jiang2013,Hua2015,He2015,Li2017,Ghane2020} have been proposed. 
Recent works favour \gls{dp}~\cite{Dwork2008} for its resistance to background knowledge attacks~\cite{Guerra-Balboa2022,Chen2020,Jiang2022}.
However, \gls{dp} introduces a privacy-utility trade-off by adding noise~\cite{Primault2014}. 
Numerous works point out that the existing \gls{dp} approaches degrade utility to a level that is no longer useful for analyses~\cite{Rao2020, Qu2020, Ma2021}.
Moreover, the added noise can lead to structural differences, enabling reconstruction attacks~\cite{RAoPT, Shao2020}.
Importantly, Miranda-Pascual et al.~\cite{Miranda-Pascual2023} uncovered flawed proofs in key \gls{dp} works like \citet{Hua2015} and \citet{Li2017}, affecting their privacy claims.
Recent approaches~\cite{Chen2020,lgan-dp} based on these mechanisms are subsequently impacted by the same issues.

Hence, we propose a design framework for privacy-preserving trajectory publication approaches based on five design goals, aiming to address and avoid the shortcomings evident in prior research.
These design goals include:
\begin{enumerate*}[label=(\roman*)]
    \item ensuring proven privacy guarantees, 
    \item defining and emphasising the \gls{uop}, 
    \item valuing practical privacy assessments alongside theoretical guarantees, 
    \item prioritising high utility and its thorough evaluation, and 
    \item considering the practical implementation of the approach.
\end{enumerate*}
Using this framework, we systematically examine existing \gls{dp} trajectory protection mechanisms, underscoring their limitations.

Acknowledging the difficulty of balancing high utility with strict privacy in (\gls{dp}) methods, a vision paper~\cite{Liu2018a} proposed \emph{trajGANs} -- Generative Adversarial Networks designed for trajectory data synthesis.
Synthetic data could replace real data in analyses, thereby safeguarding individual privacy.
Our work systematises approaches inspired by this vision, focusing on their alignment with our proposed design goals. 
LSTM-TrajGAN~\cite{Rao2020} represents a widely recognised work following this vision. 
While LSTM-TrajGAN achieves high utility with the generated data, the approach fails to provide rigorous semantic privacy guarantees.
We show the practical implications by successfully applying the attack proposed in~\cite{RAoPT} to LSTM-TrajGAN. 
Moreover, we find that no other proposed generative trajectory model achieves sufficient privacy guarantees yet.

These findings motivate an experimental study evaluating the applicability of generative models from other domains to trajectory generation.
In particular, we consider two simple \gls{rnn}-based models~\cite{Hochreiter1997}, a privacy-enhanced adaption of LSTM-TrajGAN~\cite{Rao2020}, \gls{rgan}~\cite{Esteban2017} successful for medical signal generation, WaveGAN~\cite{Donahue2019} used for audio generation, and a sequential adaptation of \gls{gpg}~\cite{GeoPointGAN} used for location generation.
First, we test the models on a toy dataset, \gls{mnist-seq}, before evaluating all models on two well-known trajectory datasets, Geolife and \gls{fs}.
While all models produce acceptable results on the toy dataset, no model can adequately capture the point distribution of the trajectory datasets. 
Despite the underwhelming performance, our findings suggest that \glsentryshort{conv1d}-based models may outperform \gls{rnn}-based ones and indicate the potential of an ensemble approach that combines point and sequential properties. 
We also identify the use of \gls{dp-sgd} as a promising method to attain genuine \gls{dp} guarantees, considering an appropriate \gls{uop} in generative trajectory models.
Our analysis concludes that designing a trajectory-generating model that offers robust privacy guarantees remains an urgent and open research gap. 
This systematisation of knowledge contributes to the field of trajectory privacy in the following ways:
\begin{itemize}
    \item We propose a framework for developing privacy-preserving trajectory publication approaches based on five pivotal design goals (\secref{sec_goals}).
    \item We contrast traditional trajectory protection methods with synthetic trajectory generation (Sections~\ref{sec_protection-types} and~\ref{sec_traditional-protection}).
    \item We critically assess the prevailing privacy-preserving trajectory synthesis techniques, identifying gaps in the current state-of-the-art (\secref{sec_trajGen}).
    \item We reproduce the results of proposed generative models and perform additional measurements, e.g., showing the success of the \gls{RAoPT} attack against LSTM-TrajGAN (\secref{sec_trajGen}).
    \item We evaluate the adaptability of six generative models to trajectory data in an experimental study, finding that none of the models is easily applicable (\secref{sec_generative-models}).
    \item We make all our experimental code available online\footnote{
    \ifanonymous
        https://github.com/ANONYMIZED
    \else
        \url{https://github.com/erik-buchholz/SoK-TrajGen}
    \fi
    }, ensuring reproducibility and facilitating future work.
\end{itemize}

\section{Background}\label{sec_background}
This section discusses the background knowledge for this work.
\secref{sec_trajectory} formally defines trajectory datasets and introduces the datasets used for evaluation.
\secref{sec_dp} explains \gls{dp}, and \secref{sec_dp-sgd} looks at \gls{dp-sgd}, the most common application of \gls{dp} to deep learning.
Then, we introduce generative models in \secref{sec_gen_models}.
Finally, \secref{sec_attacks} describes attacks against trajectory privacy methods.

\subsection{Trajectory Datasets}\label{sec_trajectory}\label{sec_datasets}

A trajectory dataset $D$ consists of a number of trajectories $D = \{T_{u1},T_{u2},\dots,T_{zm}\}$ where $T_{ui}$ refers to the $i^{th}$ trajectory of user $u$.
Each user might contribute one or multiple trajectories to the dataset.
A trajectory $T$ can be represented as an ordered sequence of locations: $T = (l_1, \dots, l_n)$.
Each location has at least two coordinates, typically latitude and longitude $l_i = (lat_i, lon_i)$, with optional additions like elevation for greater precision~\cite{Geolife1}.
Additionally, trajectories can include semantic information such as \glspl{poi} (e.g., restaurant, shop, gym) to provide context to locations~\cite{Tu2019}.
While such additional information can increase the utility of a dataset for analyses, semantic information facilitates \gls{tul}~\cite{Tu2019,marc2020}.
In a pure semantic trajectory, each location only represents a semantic location, such as a shop, without associated location information, e.g., credit card transaction datasets.
In this work, we assume each location consists of spatial coordinates, i.e., latitude and longitude.
Semantic information is optional as not all datasets record it.

\subsection{Differential Privacy}\label{sec_dp}

Privacy notions are typically categorised into two types \cite{Majeed2023}: \textit{syntactic} and \textit{semantic} notions.
\textit{\glsfirst{dp}}~\cite{Dwork2013} represents the main semantic privacy notion used to protect personal information~\cite{Guerra-Balboa2022,Majeed2023}.
The central intuition of differential privacy is that adding or removing any user's data to a dataset does not significantly change the output.
Accordingly, participation does not harm users' privacy as they have \emph{plausible deniability} regarding participation.
The mathematical definition is as follows~\cite{Dwork2013}:

\begin{definition}[Differential Privacy]
	A mechanism $\mathcal{K}$ provides $(\varepsilon, \delta)$-differential privacy if for all \textit{neighbouring} datasets $D_1$ and $D_2$, and all $S \subseteq Range(\mathcal{K})$ holds
	\begin{equation}\label{eq:dp}
		\mathds{P}[\mathcal{K}(D_1)\in S] \leq e^{\varepsilon}\times \mathds{P}[\mathcal{K}(D_2)\in S] + \delta
	\end{equation}
\end{definition}

Based on this definition, determining when two datasets are considered \textit{neighbouring} is pivotal for ensuring privacy guarantees.
Therefore, we dedicate \secref{sec_uop} to discussing the definition of \textit{neighbourhood} in the context of trajectory datasets.

Consider mechanism $\mathcal{K}$ computing a noisy average over a dataset.
If the data of another user is added to the dataset $D_1$ yielding dataset $D_2$, the change of the probabilities for the outputs of $\mathcal{K}$ is bounded through $\varepsilon$.
The smaller $\varepsilon$ is chosen, the higher the provided privacy level.
In the literature, common values for $\varepsilon$ range from \num{0.01} to \num{10}~\cite{Erlingsson2014}.
However, in machine learning using \gls{dp-sgd}, larger values for $\varepsilon$ might be used in practice~\cite{dpfyML}.
The value $\delta$ represents a failure probability for which $\varepsilon$-DP can be violated.
The best practice is to choose $\delta = \frac{1}{n}$ where $n$ is the number of records in the dataset, such that no entire record can remain unprotected~\cite{dpfyML}. 
The most common way to design a differential private mechanism is through the Laplace mechanism~\cite{Dwork2013}, Gaussian mechanism~\cite{Dwork2013}, or \gls{em}~\cite{McSherry2008}.
Differential privacy offers a few valuable properties that aid in the design of more complex mechanisms:

\subheading{Post-Processing}\label{def:post-processing}\label{sec_post-processing}
Differential privacy is \textit{immune to post-pro\-cess\-ing}~\cite{Dwork2013}.
This means that the output of any differential private mechanism can be post-processed to enhance the utility of the output without reducing the provided privacy guarantees~\cite{Dwork2013}.
However, the post-processing \textit{must not} access the original data.

\subheading{Sequential Composition}\label{sec_sequential-composition}
Moreover, the \emph{Sequential Composition Theorem} states that the sequential composition of $n$ $(\varepsilon_i,\delta_i)$-\gls{dp} mechanism provides $\sum_{i=1}^n(\varepsilon_i,\delta_i)$-\gls{dp}.
This allows designing a \gls{dp} algorithm as a combination of multiple individual algorithms.

\subheading{Parallel Composition}\label{sec_parallel-composition}
In cases where the records of a dataset are partitioned and processed by separate algorithms providing $(\varepsilon_i,\delta_i)$-\gls{dp}, such that each record is only accessed by exactly one algorithm, the combined algorithm provides $(\max_i\varepsilon_i,\max_i\delta_i)$-\gls{dp}.

\subsection{Differentially Private Deep Learning}\label{sec_dp-sgd}

Giving users or third parties access to a trained \gls{dl} model represents a common business practice \cite{DeCristofaro2020}.
Used datasets are commonly private because they contain sensitive information about individuals and business secrets, or sharing is limited by legislative constraints.
For instance, a \gls{dl}-powered text-to-image converter might be trained on a business's internal and private texts and images.
However, \glspl{mia}~\cite{Shokri2017} can recover (sensitive) training data from a released model even if the data itself is not published.
To prevent the leakage of private training data, differential privacy has been introduced to \gls{dl}~\cite{Abadi2016,dpfyML}.

We refer the reader to a recent report by \citet{dpfyML} for a thorough analysis of differentially private \gls{dl}.
Here, we briefly describe \textit{\glsfirst{dp-sgd}}~\cite{Abadi2016}, which represents the most widely implemented application of \gls{dp} to deep learning~\cite{dpfyML}.
The first step of \gls{dp-sgd} is clipping the gradients to a clipping norm $C$ such that the influence of each training sample is bounded.
Second, Gaussian noise is added to the gradients based on this clipping norm $C$ and a noise multiplier $\sigma$.
Third, privacy accounting tracks the $\varepsilon$ budget during training to ensure a $(\varepsilon, \delta)$-\gls{dp} guarantee upon completion.

\subsection{Generative Models}\label{sec_gen_models}
Generative models for sequential data~\cite{Eigenschink2023}, such as trajectories, have been proposed to generate more data for domains with limited data available or to generate synthetic data in order to replace real data that is sensitive, e.g., in the medical domain~\cite{Esteban2017}.
Here, we focus on those architectures used in the context of trajectories.

\glsreset{rnn}
\subheading{\gls{rnn}}
\textit{\glspl{rnn}} sequentially process inputs by maintaining a memory of previous inputs through a hidden state that is passed on from one cell to the next.
This makes \glspl{rnn} well-suited for trajectories, as each location relies on previous ones.
The most common type of \gls{rnn} in the context of trajectories is the \gls{lstm}~\cite{Hochreiter1997}, which improves the ability to understand long-term dependencies.

\subheading{\gls{ae}}
\textit{\glspl{ae}}~\cite{autoencoders} consist of an encoder that encodes a given input into a lower-dimensional latent representation and a decoder that aims at reconstructing the original input.
Common applications are dimensionality reduction and denoising.
The latent bottleneck forces outputs to be different, while the most important information is preserved if the model is trained correctly.

\subheading{\gls{vae}}
\textit{\glspl{vae}}~\cite{vae2013} are a probabilistic extension of \glspl{ae}.
The encoder maps inputs to latent \textit{distributions}, enforced by minimising the KL divergence to the target distribution.
The decoder then samples from the latent space distributions to generate data.
This enables \glspl{vae} to both reconstruct and generate.

\subheading{\gls{gan}}\label{sec_gan}
\textit{\glspl{gan}}~\cite{Goodfellow2014} are composed of two separate models, the generator $G$ and the discriminator $D$.
The generator $G$ receives noise as input and tries to generate samples that are indistinguishable from real data samples.
Meanwhile, the discriminator $D$ judges whether a given input sample comes from the real dataset or the generator.
The resulting zero-sum game is represented by the adversarial loss~\cite{Goodfellow2014}:
\begin{equation}
\min_G \max_D \mathbb{E}_{x\sim p_{\text{data}}}[\log D(x)] + \mathbb{E}_{z\sim p_z}[\log(1 - D(G(z)))]
\end{equation}
Here, $p_{\text{data}}$ is the real data distribution, and $p_z$ is $G$'s input noise distribution.
The generator and discriminator train iteratively until $G$ produces samples that can barely be distinguished from real data.

\subheading{\gls{aae}}
\textit{\glspl{aae}}~\cite{aae2016} combine \glspl{ae} with the adversarial concept of \glspl{gan}.
The encoder transforms inputs into latent representations that approximate a target distribution.
In contrast to \glspl{vae}, \glspl{aae} employ an adversarial loss, i.e., a discriminator's feedback, to ensure the latent space adheres to the designated distribution.
Like \glspl{vae}, \glspl{aae} can reconstruct input data and generate new synthetic data.

\subsection{Practical Attacks against Trajectory Privacy}\label{sec_attacks}
Several attacks on trajectory protection highlight the risks of insufficient protection.
\citet{Miranda-Pascual2023} provide a comprehensive overview. 
Here, we focus on two practical attacks, namely \gls{RAoPT}~\cite{RAoPT} and \gls{tul}~\cite{marc2020}, which provide open-source code~\cite{Buchholz2022Code, marc2020_code} such that we propose them for empirical evaluation in \secref{goal_practical}.

\subheading{\gls{tul}}
While a single trajectory might not reveal sensitive information or allow identification, linking multiple trajectories to the same user increases these risks~\cite {Rao2020, marc2020}. 
Therefore, some protection mechanisms~\cite{Rao2020, Shin2023, Song2023, Fontana2023} utilise \textit{\gls{tul}} to evaluate the provided privacy level, with a lower \gls{tul} success rate implying better privacy.
One recent and openly available~\cite{marc2020_code} \gls{tul} algorithm is MARC~\cite{marc2020}.
This \gls{dl} model consisting of embedding, \gls{rnn}, and softmax layers matches multi-aspect trajectories containing semantic information, e.g., \gls{poi} information, to users.

\subheading{\gls{RAoPT}}
\citet{RAoPT} note that the perturbation of \gls{dp} protection mechanisms causes structural differences in the outputs, e.g., zigzag patterns.
They develop the \gls{RAoPT} \gls{dl} model, which trains on protected samples as inputs and real samples as targets to understand their relationship.
The trained model can then reconstruct versions closer to the originals from unseen protected trajectories, thereby reducing perturbations and compromising privacy.

\section{Design Goals}\label{sec_goals}

This section introduces our design framework for privacy-pre\-serv\-ing trajectory publication mechanisms.
The overarching goal can be defined as:
\textit{The release of high-utility trajectory data without revealing private information about the individuals participating in the dataset.}
In the following, we discuss the five goals forming our design framework.
These goals are \textit{not} ordered by importance, as this depends on the specific dataset and use case.

\goal{Formal Privacy Guarantees}{goal_guarantees}

Ensuring stringent privacy guarantees is essential when releasing trajectory datasets. 
While practical privacy evaluations against specific attacks have advantages, as we discuss in \secref{goal_practical}, formal guarantees are irreplaceable.
First, testing against a designated attack only demonstrates resistance to that specific threat.
It does not vouch for protection against other existing attacks, unforeseen future threats, or enhanced algorithms of the same attack.
Second, this method lacks quantifiable privacy guarantees.
In a commercial context, evidence of rigorous efforts to safeguard shared data is necessary.
Leveraging formal guarantees, like \gls{dp} with established $\varepsilon$ and $\delta$ parameters, provides such evidence.
It signifies that state-of-the-art protection, endorsed by the research community at the time of release, was implemented.
In case of a privacy incident, these guarantees testify to the releasing entity's earnest attempts at privacy preservation.
Differential privacy stands out as the de facto standard for semantic privacy guarantees.
However, specific scenarios might allow for \gls{dp} relaxations.
One notable example is \gls{lldp} used by \gls{gpg} (\refer \secref{sec_pointgen}).

\subheading{\gls{ldp}}
In standard (global) \gls{dp}, a central entity applies protections, such as noise addition, to data collected from clients before sharing it.
This requires trust from the clients in the central entity. 
\gls{ldp} eliminates this trust requirement by applying protection \textit{locally} on the client's device before sending it to the central entity, enhancing privacy guarantees at the cost of potentially greater utility loss.
This work presumes trust in the central data collector as global \gls{dp} is the most common notion for trajectory privacy.
However, we highlight when approaches provide the stronger \gls{ldp} guarantees.

\subheading{Common Pitfalls}
\citet{Miranda-Pascual2023} demonstrate that multiple foundational works on \gls{dp} trajectory protection rely on erroneous proofs, making quantifying the provided privacy level challenging.
We highlight similar issues affecting further works in Sections~\ref{sec_protection} and~\ref{sec_trajGen}.
While the formal proofs for these flaws are provided in \cite{Miranda-Pascual2023}, we summarise the most common pitfalls:
\begin{enumerate}[nosep, wide, labelwidth=0pt, labelindent=0pt]
    \item \textbf{Not Considering All Elements~\cite{Miranda-Pascual2023}:}\label{pitfall:domain}
    A \gls{dp} mechanism must be able to output any \textit{possible} value \textit{independent of the dataset}~\cite{Miranda-Pascual2023}.
    However, some methods~\cite{Zhao2019, Zhao2020, Zhao2021, Yuan2021, lgan-dp} only add noise to existing elements, leaving unobserved elements with a count of $0$.
    Consequently, an element $e$ not present in dataset $D$ would always have a count of $0$.
    If neighbouring dataset $D'$ contains $e$, the difference between $D$ and $D'$ cannot be bounded by $\varepsilon$ (\refer \hyperref[eq:dp]{Equation~\ref*{eq:dp}}).
    This contradicts the definition of $\varepsilon$-\gls{dp}.
    \glsreset{em}
    \item \textbf{\gls{em} Application~\cite{Miranda-Pascual2023}:}\label{pitfall:em}
    The \gls{em}'s outputs depend on a score function $u(d, r): D \times R \rightarrow \mathbb{R}$ that assign a score $u$ to each possible output $r\in R$ based on the dataset $d \in D$.
    However, the set of all possible datasets $D$ and outputs $R$ must be well-defined.
    Several studies~\cite{Hua2015,Li2017,Chen2020,Ma2021,Zhang2023,Haydari2022} inaccurately define $R$ as dependent on the selected dataset $d$.
    This renders the \gls{em} indefinable and invalidates any \gls{dp} claim based on it~\cite{Miranda-Pascual2023}.
    \item \textbf{Access to Real Data:}\label{pitfall:data-access}
    The post-processing property permits any manipulation of \gls{dp} outputs \textit{not accessing the original data} (\refer\secref{sec_dp}).
    However, certain methods~\cite{dp-trajgan} access the raw data via side channels, undermining any \gls{dp} protection measures.
\end{enumerate}

\subheading{Conclusion}
Semantic privacy guarantees are inevitable even in the presence of privacy evaluations.
The gold standard is \glsentrylong{dp}, but relaxations might be justifiable.
\gls{ldp} provides stronger guarantees, making it a preferable choice when achievable.
Rigorous verification of privacy claims is essential, highlighted by some baseline studies' flawed \gls{dp} proofs.

\goal{Unit of Privacy}{goal:uop}\label{sec_uop}

For privacy guarantees, choosing the correct unit of data to protect is of utmost importance.
The choice of a \textit{\glsfirst{uop}} is equivalent to the question of what makes two inputs (datasets) of a \gls{dp} mechanism ``neighbouring`` (\refer \secref{sec_dp}).
A larger \glsentryshort{uop} requires more obfuscation to achieve the same privacy level.
Thus, the \gls{uop} should be chosen as small as possible yet large enough to ensure privacy.
Protecting a too-small unit of data is a common problem and can lead to vulnerability to correlation~\cite{Miranda-Pascual2023} or reconstruction attacks~\cite{RAoPT}.
Recent literature discussed the \gls{uop}, e.g., \citet{dpfyML} focusing on machine learning and \citet{Miranda-Pascual2023} (using the term \textit{level of granularity}) for trajectory data.
We use the term \gls{uop} and provide a consolidated systematisation specific to location trajectories.
This shall support researchers in selecting the appropriate \gls{uop} for their application.
In the following, we consider the dataset $D_1$ consisting of multiple trajectories $D_1 = (T_{u1},T_{u2}\dots,T_{zm})$, where $T_{ui}$ is the $i^{th}$ trajectory of user $u$.

\noindent\textbf{User-level Privacy}
    corresponds to the original definition of differential privacy~\cite{Dwork2013}.
    For a trajectory dataset $D_1$, $D_2$ can be obtained by removing all trajectories belonging to one user $u$: $D_2 = D_1 \setminus \{T_{vi} | v = u\}$.
    While user-level privacy offers the most robust protection, it often leads to a notable reduction in data utility.
    As user-level privacy represents the original definition of \gls{dp}, all of the following units of privacy constitute relaxations thereof.

\noindent\textbf{Instance-Level Privacy},
    or \textit{example-/trajectory-level} privacy, safeguards individual trajectories $T_{ui}$ with a specific privacy guarantee.
    According to the composition theorem (\refer \secref{sec_dp}), $\varepsilon$ instance-level differential privacy yields $m\varepsilon$ user-level differential privacy for a user contributing $m$ trajectories. 
    Hence, instance-level and user-level privacy are equivalent for datasets containing one trajectory per user.
    Due to the protection of a trajectory as one unit, this level protects against any attack exploiting the correlation within a trajectory.
    Instance-level \gls{dp} is the most common level in \gls{dl}~\cite{dpfyML}, as \gls{dp-sgd} provides instance-level privacy for the training samples.

\noindent\textbf{Location-Level Privacy}
    refers to \textit{event-level} privacy in the general setting~\cite{Miranda-Pascual2023}.
    In the case of a trajectory $T = (l_1,\dots,l_n)$, each location $l_i$ is protected with the privacy guarantee independent of the other locations.
    This weakest level of privacy requires the least perturbation. 
    While literature~\cite{Kim2022, Jiang2013} uses location-level privacy for trajectory protection, it is vulnerable to correlation and reconstruction attacks.
    Early works on location privacy~\cite{Andres2013} already warned against using such solutions for trajectory projection.

\noindent\textbf{Multi-event-level Privacy}
    bridges instance- and location-level privacy, accommodating notions such as $w$-event-level and $l$-tra\-jec\-to\-ry-level privacy (see \cite{Miranda-Pascual2023} for a detailed discussion).
    It covers a window of multiple events within a trajectory, equating to location-level privacy at $w=1$ and instance-level at $w=\max\{|T_i| \in D\}$.
    This approach is useful when sensitive information, like a hospital visit, spans multiple locations such that location-level privacy is insufficient, yet trajectory-level privacy is too restrictive.

\subheading{Example}
To demonstrate the significance of selecting the appropriate \gls{uop}, consider the trajectory protection methods CNoise and the \glsentryshort{sdd} mechanism \cite{Jiang2013}.
Despite their \gls{dp} guarantees, reconstruction attacks \cite{Shao2020, RAoPT} successfully target these mechanisms.
These attacks do not contradict the rigorous definition \gls{dp}.
Instead, both attacks leverage the correlations between and within trajectories.
The wrong \gls{uop}, i.e., protecting individual locations instead of the entire trajectory, makes these mechanisms vulnerable.
Nevertheless, recent works~\cite{Kim2022} still use location-level approaches in the context of trajectory privacy (\refer \secref{sec_trajGen}).

\subheading{Conclusion}
Given these findings, we strongly advise researchers to carefully select the \gls{uop} and refrain from using location-level privacy for trajectories.
We view instance-level privacy as a promising balance between utility and privacy due to the protection against intra-trajectory correlations and its applicability to \gls{dl}.
However, employing instance-level privacy requires special care for recurring locations in multiple trajectories, such as home or work.

\goal{Empirical Privacy}{goal_practical}

Proof of formal privacy guarantees (\grefshort{goal_guarantees}) and adherence to the appropriate \gls{uop} (\grefshort{goal:uop}) might make a practical privacy assessment seem unnecessary.
However, as \citet{Miranda-Pascual2023} showed, numerous \gls{dp} trajectory protection mechanisms are built on flawed \gls{dp} proofs.
Additionally, vulnerabilities can arise from the incorrect selection of the \gls{uop}, as elaborated in \secref{sec_uop}.
Given these common challenges, we advocate for integrating practical privacy evaluations into future publications to identify potential issues and emphasise the achieved privacy levels.

Recent studies~\cite{Rao2020, Song2023, Shin2023, Fontana2023} have employed the publicly available \textit{\glsfirst{tul} algorithm MARC}~\cite{marc2020,marc2020_code} as practical privacy assessment.
Additionally, the novel \textit{reconstruction attack \gls{RAoPT}~\cite{RAoPT}} provides publicly available code~\cite{Buchholz2022Code},  making it a suitable evaluation tool for trajectory protection mechanisms.
Several approaches \cite{Li2017,Chen2020,Liu2021a,dp-trajgan,lgan-dp,Zhang2023}
use the \textit{\gls{mi}} metric to measure the provided privacy level.
\gls{mi} measures the dependence of the protected dataset on the original dataset such that a lower value indicates better privacy.
However, as the goal for the published dataset is to retain as much useful information as possible while removing all private information, a low \gls{mi} might also indicate low utility.
Therefore, we consider the aforementioned practical attacks more meaningful in quantifying practical privacy.

\subheading{Conclusion}
While not strictly necessary, we recommend evaluating trajectory publication approaches against at least one practical attack, even with formal privacy guarantees.
The open-source attacks \gls{tul}~\cite{marc2020_code}, and \gls{RAoPT}~\cite{Buchholz2022Code} represent suitable candidates.

\begin{table}[t]
    \begin{tabular}{|c|c|c|}
    \toprule
     & \textbf{Preservation} & \textbf{Statistics} \\
     \midrule
    \textbf{Point} & \glsentryshort{hd} & Range Query \\
    \textbf{Level}  &   Density & Hotspot Preservation \\
     \midrule
    \textbf{Trajectory} & \glsentryshort{dtw}$\dagger$ & Travelled Distance \\
    \textbf{Level} & \glsentryshort{hd}$\dagger$ & Segment Length \\
     \bottomrule
    \end{tabular}
    \caption{
    Utility Metric Systematisation.
    Metrics with $\dagger$ require a 1:1-mapping between input and output trajectories.
    }
    \label{tab_metrics}
    \vspace{-1.8em}
\end{table}

\goal{Utility}{goal_utility}

The primary aim of a released dataset is the usability for downstream tasks and analyses.
Any protection mechanism has to trade off privacy and utility~\cite{Primault2014}.
Thus, utility represents our fourth design goal.
We systematise utility metrics, discuss data aggregation and precision, and address environmental constraints.

\subheading{Utility Metrics}\label{sec_utility-metrics}
Recent studies \cite{Miranda-Pascual2023, Su2020} have analysed various trajectory utility metrics.
Rather than reiterating these, we aim to establish a selection framework. 
The large number of metrics highlights the core issue.
The absence of a universally adopted metric makes comparing different methodologies difficult.
Thus, we recommend using established metrics in future research.

We propose classifying utility metrics along two axes as shown in \tabref{tab_metrics}.
First, metrics can be separated into \textit{data preservation} and \textit{statistical} metrics~\cite{Miranda-Pascual2023}. 
Data preservation metrics evaluate original data retention, while statistical metrics assess the conservation of overall patterns. 
Second, metrics can be oriented towards \textit{point properties} or \textit{sequential properties}.
A prevailing trend is emphasising point-level metrics, potentially neglecting sequential coherence.

\begin{figure}
    \centering
    \includegraphics[width=\linewidth]{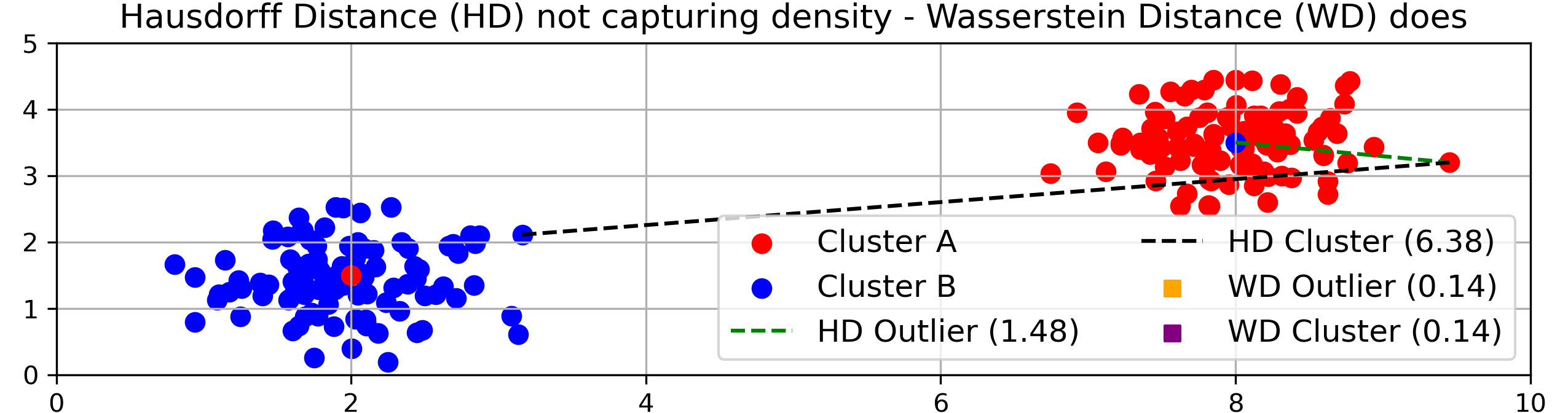}
    \vspace{-0.6cm}
    \caption{
    Density problem of \gls{hd}.
    A single outlier significantly reduces the \glsentryshort{hd} but not the \glsentryshort{wd}.
    }
    \label{fig_hd-issue}
\end{figure}

We present example metrics covering all categories, focusing on continuous, spatial coordinate trajectories.
For a comprehensive list, see \cite{Miranda-Pascual2023, Su2020}.
Metrics concentrate on spatial dimensions as minimal information of trajectories (\refer \secref{sec_trajectory}).
Datasets with additional properties, like semantics, should incorporate these, e.g., via the \textit{\gls{stc}} \cite{Miranda-Pascual2023, Cunningham2021-DPGEN}.
For \textit{point preservation}, the \textit{\glsfirst{hd}} is commonly used \cite{Hua2015, Chen2020, Liu2021a}, comparing the distance of points in one dataset to their nearest counterparts in another.
However, it may not effectively capture discrepancies in point distributions' densities.
Consider the two clusters $A$ (red) and $B$ (blue) in \figref{fig_hd-issue}.
First, assume that the clusters are distinct, i.e., ignore the single point with the opposite cluster. 
Then, the black line represents the \gls{hd} with the value $6.38$. 
However, with a single outlier of $B$ in the centre of $A$, the \gls{hd} reduces to the green line, resulting in a value of $1.48$.
This sensitivity to outliers limits the \gls{hd}'s effectiveness for comparing generated ($B$) and real ($A$) point clouds.
Therefore, we suggest additionally comparing the spatial distributions using metrics like the \gls{wd} or \gls{jsd}.
In the example, the value of the \gls{wd} remains consistent at $0.14$.
In terms of \textit{point statistics}, \textit{range queries} \cite{Hua2015,Gursoy2018,Liu2021a,Cunningham2021-LDP,Cunningham2021-DPGEN,GeoPointGAN} and its discrete counterpart count queries \cite{Chen2011,N-grams2012,Haydari2022,Sun2023} have gained prominence, often paired with \textit{hotspot preservation queries} \cite{Gursoy2018,Cunningham2021-LDP,Cunningham2021-DPGEN,GeoPointGAN,TSG}.
\textit{Trajectory preservation} metrics usually require a \textit{1:1-mapping} for comparison, i.e., each output trajectory should match an original trajectory. 
However, some generative models, such as \glspl{gan}, create new samples from noise input (\refer\secref{sec_gan}) without a clear link to an original sample.
This makes it unclear which trajectories to match for comparison.
Hence, applying trajectory preservation metrics to some generative models is difficult.
One possible solution is matching each generated trajectory to the closest original.
If a 1:1 mapping is given, the spatial metrics \gls{hd} \cite{Rao2020,Liu2021a,Ma2021,Kim2022,Jiang2023} and \gls{dtw} \cite{Jiang2013, Jiang2023, Zhang2024} represent suitable candidates.
Note that in this case, the \gls{hd} between trajectories are computed, whereas, for point preservation, the \gls{hd} between (an equal-sized subset of) all points of the original and protected dataset is evaluated.
We note a research gap in trajectory preservation metrics for generative models without 1:1 mapping.
Lastly, \textit{Trajectory statistics} should focus on the sequential properties. 
Therefore, we deem the total \textit{Travelled distance} \cite{He2015,Gursoy2018,Ghane2020,TSG,Haydari2022,Jiang2023} and the \textit{Segment lengths} \cite{He2015,Gursoy2018} as appropriate metrics.
The former computes the total distance travelled from start to end point for each trajectory, while the latter computes the distance between any two consecutive points. 
Then, the resulting distributions of the original and generated dataset are compared, e.g., via \gls{jsd} or \gls{wd}.
However, metrics are not the only utility evaluation factor.

\subheading{Aggregation}\label{sec_aggregation}
Various \gls{dp} methods release aggregated datasets, presenting counts for representative trajectories \cite{Hua2015, Li2017, Chen2020, Liu2021a}.
This approach requires less location obfuscation since the information from multiple trajectories can be consolidated.
However, aggregation invariably results in detail loss.
Thus, individual trajectory release is preferable for maximal utility, especially if the data's eventual application is unknown at release time.

\subheading{Precision}\label{sec_precision}\label{coordinate-trajectory}
Utility in data preservation is directly linked to precision.
Broadly, trajectories can be categorised into \textit{\gls{poi}-trajectories} and \textit{coordinate trajectories}.
The former involves locations from a predefined, countable set of \glspl{poi}.
In contrast, the latter typically involves arbitrary coordinates, like latitude and longitude.
Generally, \gls{poi}-based methods yield higher utility at a given privacy level than coordinate-based ones.
However, this article focuses on coordinate trajectories due to their broader applicability.

\subheading{Grid Approaches}\label{sec_grid-utility}
Some solutions \cite{He2015, Ghane2020} convert the continuous space into a grid, discretising the data by assigning each location to a grid cell.
While this shares some benefits with \gls{poi} trajectories, it introduces challenges.
The precision is constrained by grid granularity \cite{TSG}, which is limited by computational considerations.
A finer grid translates into a more extensive range of values, increasing the computational effort.
Furthermore, the grid's spatial extent is typically limited (e.g., to a city) to maintain meaningful cell sizes.

\subheading{Environmental Constraints}\label{sec_environment}
Recent literature \cite{Naghizade2020,RAoPT,GeoPointGAN,Cunningham2021-LDP,Cunningham2021-DPGEN,Miranda-Pascual2023,Sun2023} underlines the importance of \textit{environmental} constraints.
Protection methods can cause violations such that obfuscated trajectories may pass through rivers or physical barriers \cite{RAoPT, Cunningham2021-DPGEN}, consecutive locations may not be reachable \cite{Cunningham2021-LDP}, or cars do not follow the road network \cite{RAoPT}.
Considering these constraints improves utility and prevents using this knowledge for reconstruction attacks \cite{RAoPT}. 

\subheading{Conclusion}
For meaningful comparisons among approaches, it is vital to employ \textit{established utility metrics}.
These metrics should cover both \textit{point} and \textit{trajectory aspects} and ideally evaluate both \textit{data preservation} and \textit{statistical attributes}.
The release of \textit{individual trajectories} is preferable over aggregated data.
Solutions for \textit{coordinate trajectories} are most generally applicable, but \textit{\gls{poi} datasets} or \textit{grid-based} approaches might present optimised results for some cases.
Finally, considering \textit{environmental constraints} is essential.

\goal{Practicality}{goal_feasibility}

Releasing a dataset usually occurs once, making computational cost a secondary concern to utility and privacy.
However, \textit{practicality} remains crucial, i.e., protection approaches must not require high-cost, specialised equipment but commodity hardware should suffice.
Furthermore, a predictable runtime is preferable.
For example, the \gls{sdd} mechanism's~\cite {Jiang2013} probabilistic sampling can yield uncertain and extended runtimes of many hours as confirmed by the \gls{RAoPT} artefacts~\cite{Buchholz2022Code}.
Moreover, evaluations should consider at least two datasets, ideally publicly available.
Trajectory datasets differ significantly in granularity (\SI{5}{\second} vs. per minute), span (city-level vs. global), and additional information.
Beyond the challenges of differing datasets and metrics, reproducibility remains a hurdle, often due to insufficient publication details and unresponsive authors. 
LSTM-TrajGAN~\cite{Rao2020} demonstrates the benefits of publishing source code (\refer \secref{sec_trajGen}).
While LSTM-TrajGAN influences many generative trajectory models, models without public code often receive limited attention.

\subheading{Conclusion}
The trajectory publication approach should 
\begin{enumerate*}[label=(\roman*)]
    \item run on \textit{commodity hardware}, 
    \item have a \textit{deterministic runtime}, 
    \item be evaluated on \textit{at least two public datasets}, and 
    \item ideally \textit{share the source code} for reproducibility.
\end{enumerate*}
Based on these five goals, we examine existing trajectory publication schemes in the following.

\begin{figure}
    \centering
    \begin{tikzpicture}[
        font=\small,
        level distance=0.75cm,
        level 1/.style={sibling distance=3cm},
        level 2/.style={sibling distance=2cm},
        level 3/.style={sibling distance=2cm}]
        
    \node (root) {Trajectory Publication}
        child {node {\glsentrylong{dp}}
            child {node {Protection}
                child {node {\secref{sec_protection}}}
            }
            child {node[align=center,xshift=1cm] (dl) {Deep learning-\\based Generation}
                child {node[] {\secref{sec_trajGen}}}
                child {node[] {\secref{sec_generative-models}}}
            }
        }
        child {node[xshift=0.2cm] {Syntactic Privacy}};

    \draw[dashed] (root) -- (dl);
    
    \end{tikzpicture}
    \vspace{-0.5em}
    \caption{Categorisation of trajectory publication.}
    \label{fig_trajectory_publication}
    \ifanonymous
    \else
        \vspace{-2em}
    \fi
\end{figure}
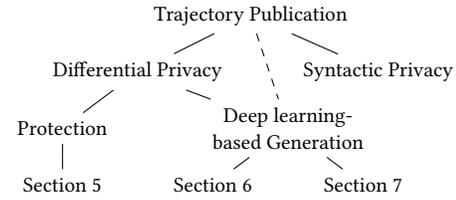

\section{Systematisation of Approaches}\label{sec_protection-types}

We outlined our design framework for trajectory publication mechanisms in the preceding section.
Such mechanisms are classified based on several attributes, highlighted in \figref{fig_trajectory_publication}.
The primary distinction hinges on the targeted privacy notion, typically categorised into \textit{syntactic} and \textit{semantic} notions~\cite{Majeed2023}.
\textit{Syntactic privacy notions}, such as $k$-Anonymity \cite{Sweeney2002}, though common, are susceptible to background knowledge attacks and do not provide rigorous guarantees \cite{Andres2013, Chen2020, lgan-dp, dp-trajgan}.
Conversely, \gls{dp} (\refer \secref{sec_dp}), the leading \textit{semantic privacy notion}, represents the current de facto privacy standard due to its strong guarantees.
Furthermore, some methods, like \glspl{gan}-based ones, ensure privacy by generating new data rather than targeting specific privacy notions. 
Yet, these mechanisms may also incorporate privacy notions alongside their generative properties.

This work discusses \textit{\gls{dp} protection mechanisms} (\secref{sec_protection}) and \textit{\glsfirst{dl}-based trajectory generation} (\secref{sec_trajGen} and~\ref{sec_generative-models}).
To explore k-anonymity-based methods, readers are referred to \cite{Jin2021}.
The separation between protection and generation approaches overlaps, as some obfuscation-based protections, such as DPT~\cite{He2015}, record noisy statistics of the original dataset and sample from these to generate protected trajectories.
We achieve clarification by categorising approaches based on \gls{dl} as \textit{\gls{dl}-based trajectory generation} while we discuss all other approaches jointly as \textit{\gls{dp} protection mechanisms}.
The DL-based trajectory generation approaches represent a recent development, with the first vision proposed in 2018~\cite{Liu2018a}.
To the best of our knowledge, they have not been thoroughly discussed and systemised regarding privacy preservation and represent the focus of this work.
Next, we discuss the protection mechanisms and describe the state-of-the-art trajectory generation in \secref{sec_trajGen}.

\section{Differential Privacy Protection}\label{sec_traditional-protection}\label{sec_protection}

\begin{table*}
    \centering
    \begin{tabular}{l l c c c c c c c p{5cm}}
        \toprule
        \textbf{Category} & \textbf{Approach(es)} & \textbf{In~\cite{Miranda-Pascual2023}} & \textbf{\gls{uop}} &  \textbf{\ref{goal_guarantees}} &  \textbf{\ref{goal:uop}} &  \textbf{\ref{goal_practical}} &  \textbf{\ref{goal_utility}} & \textbf{\ref{goal_feasibility}} & \textbf{Main Shortcoming} \\
        \midrule
        \ballnumber{A} Tree-based &\cite{Chen2011,N-grams2012} & $\checkmark$ & Instance &   & $\checkmark$ & $-$ & \xmark & \xmark & Requires pre-defined location domain \\ \hline
        \multirow{2}{*}{\ballnumber{B} Noisy Count} & \multirow{2}{*}{\cite{Zhao2019,Zhao2020,Zhao2021,Yuan2021}} & \multirow{2}{*}{$\checkmark$} & \multirow{2}{*}{Instance}  & \multirow{2}{*}{\xmark} & \multirow{2}{*}{$\checkmark$} & \multirow{2}{*}{$-$} & \multirow{2}{*}{\xmark} & \multirow{2}{*}{$\checkmark$} & Broken \gls{dp} proof: Only existing traj. considered (\pitfall{domain})\\ \hline
        \multirow{4}{*}{\ballnumber{C} Clustering Based} & \multirow{2}{*}{\cite{Hua2015,Li2017,Chen2020}} & \multirow{2}{*}{$\checkmark$} & \multirow{2}{*}{Instance}  & \multirow{2}{*}{\xmark} & \multirow{2}{*}{$\checkmark$} &\multirow{2}{*}{$-$} & \multirow{2}{*}{\xmark} & \multirow{2}{*}{$\circ$} & Broken \gls{dp} proof (\gls{em} -- \pitfall{em}); Aggregated output \\
        &\cite{Ma2021,Zhang2023} & \multirow{2}{*}{\xmark} & \multirow{2}{*}{Instance}  & \multirow{2}{*}{\xmark} & \multirow{2}{*}{$\checkmark$} &\multirow{2}{*}{$-$} & \multirow{2}{*}{\xmark} & \multirow{2}{*}{$\circ$} & Broken \gls{dp} proof (\gls{em} -- \pitfall{em}); Aggregated output \\\hline
        \multirow{2}{*}{\ballnumber{D} Location Privacy} & \multirow{2}{*}{\cite{Jiang2013, Andres2013, Chatzikokolakis2014}} & \multirow{2}{*}{\xmark} & \multirow{2}{*}{Location}  &   & \multirow{2}{*}{\xmark} & \multirow{2}{*}{\xmark} & \multirow{2}{*}{$\circ$} & \multirow{2}{*}{$\circ$} & Wrong \gls{uop}; Successful attack known for~\cite{Jiang2013}; Length-based utility decline \\ \hline
        \multirow{2}{*}{\ballnumber{E} Grid-Cell based} & \multirow{2}{*}{\cite{He2015,Ghane2020,Sun2023}} & \multirow{2}{*}{\xmark} & \multirow{2}{*}{Instance}  &   & \multirow{2}{*}{$\checkmark$} & \multirow{2}{*}{$-$} & \multirow{2}{*}{$\circ$} & \multirow{2}{*}{$\circ$} & Grid cell granularity limits precision or performance \\ \hline
        \multirow{2}{*}{\ballnumber{F} \parbox{2cm}{ Environmental\\Constraints}} &\cite{Haydari2022} & \xmark & Instance  & \xmark & $\checkmark$ & $-$ & $\checkmark$ & $\checkmark$ & Broken \gls{dp} Proof (\gls{em} -- \pitfall{em}) \\
        &\cite{Cunningham2021-DPGEN} & \xmark & Location  &   & \xmark & $-$ & $\checkmark$ & $\checkmark$ & Generates points only \\ \hline
        \multirow{2}{*}{\ballnumber{G} \gls{poi} protection} & \multirow{2}{*}{\cite{Cunningham2021-LDP}} & \multirow{2}{*}{$\checkmark$} & \multirow{2}{*}{Instance}  &   & \multirow{2}{*}{$\checkmark$} & \multirow{2}{*}{$-$} & \multirow{2}{*}{\xmark} & \multirow{2}{*}{$\checkmark$} & Only for \gls{poi} data; Requires public knowledge to be available \\
        \bottomrule
    \end{tabular}
    \caption{
    Overview of Differentially Private Protection Mechanisms discussed in \secref{sec_traditional-protection}.
    The table lists each approach's inclusion in the SoK~\cite{Miranda-Pascual2023}, targeted \gls{uop}, adherence to design goals from \secref{sec_goals}, and primary shortcomings.
    Symbols: $\checkmark$~(goal satisfied), $\circ$~(partially satisfied), and \xmark~(unsatisfied).
    For \textbf{\ref{goal_guarantees}}, \xmark denotes known \gls{dp} proof errors; the absence of a symbol means we are unaware of any flaw.
    For \textbf{\ref{goal_practical}}, \xmark indicates demonstrated attacks and approaches with $-$ have not been evaluated.
    }
    \label{tab_trajectory-protection}
    \vspace{-1.5em}
\end{table*}

\citet{Chen2011} first introduced the application of \glsfirst{dp} to location trajectories in 2011. 
Since this pioneering work, various approaches have evolved in the \gls{dp} trajectory protection field.
This Section briefly addresses these protection methods, highlighting their limitations in meeting the design goals outlined in \secref{sec_goals}.
These limitations motivate alternative solutions, such as \gls{dl}-based approaches for trajectory generation (\secref{sec_trajGen}), which represent the focus of this work.
\citet{Miranda-Pascual2023} provide a comprehensive analysis of \gls{dp} trajectory protection mechanisms in their SoK. 
\tabref{tab_trajectory-protection} summarises our observations and indicates the approaches detailed by \citet{Miranda-Pascual2023}.

\noindent\ballnumber{A}\subheading{Tree-based Approaches}
The first \gls{dp} mechanism~\cite{Chen2011} is based on noisy prefix trees, i.e., a tree representing all possible sequences of locations.
The occurrences of each prefix are counted, and Laplace noise is added to achieve \gls{dp} before reconstructing output trajectories based on the noisy tree.
The authors improved the approach's utility by using n-grams in subsequent work~\cite{N-grams2012}.
However, all mechanisms in this line of research rely on a pre-defined set of discrete locations. 
Hence, these approaches only apply to \gls{poi} datasets but not to general coordinate trajectories, violating \gref{goal_utility}.
Enumerating all locations within a particular area, e.g., using a grid, incurs unreasonable computational overhead as the tree size increases exponentially~\cite{Miranda-Pascual2023}, violating \gref{goal_feasibility}.

\noindent\ballnumber{B}\subheading{Noisy Count-based Approaches}
Other approaches~\cite{Zhao2019,Zhao2020,Zhao2021,Yuan2021} utilising noisy counts are summarised in~\cite{Miranda-Pascual2023}.
The authors show that these methods contain an error in their \gls{dp} proofs, violating \gref{goal_guarantees}.
Only the counts for existing sequences are perturbed, but not for all possible (count zero) sequences (\pitfall{domain}).
However, a \gls{dp} mechanism has to produce any output with a probability larger than zero.
Otherwise, the mechanism's outputs cannot be bounded (ref.~\cite{Miranda-Pascual2023} for a formal proof).

\noindent\ballnumber{C}\subheading{Clustering-based Approaches}
One line of research~\cite{Hua2015, Li2017, Chen2020, Ma2021, Zhang2023} is based on clustering the locations at timestamps, followed by sampling from these clusters.
\citet{Miranda-Pascual2023} show that the \gls{em} is used incorrectly by these approaches (\pitfall{em}), breaking the formal \gls{dp} proof:
Proper \gls{em} application requires dataset-independent output sets, whereas these methods use the dataset-dependent set of possible partitions, breaching \gref{goal_guarantees}.
Recent studies~\cite{Ma2021, Zhang2023}, not covered in \cite{Miranda-Pascual2023}, repeat this error.
Furthermore, they aggregate data into reference trajectories with noisy counts, thereby reducing utility (\gref{goal_utility}).
They also risk generating implausible trajectories due to poor cluster representation choices and assume uniform lengths and sampling rates, which is impractical for real-world data~\cite{Miranda-Pascual2023}.

\noindent\ballnumber{D}\subheading{Location-level Privacy}\label{sec_location-privacy}
Other approaches protect each trajectory location individually, such as the \gls{sdd} mechanism and CNoise~\cite{Jiang2013}.
The former samples a distance and direction from one location to the next, while the other adds Laplace noise to each location.
These mechanisms violate \gref{goal:uop}, as the privacy budget $\varepsilon$ is used for a single location.
This makes the protection vulnerable to reconstruction attacks like \gls{RAoPT}~\cite{RAoPT} (violating \gref{goal_practical}).
Moreover, the \gls{sdd} mechanism requires probabilistic sampling until a suitable next location is found, which can lead to run times of several hours for datasets such as Geolife~\cite{Geolife1} (see Algorithm~5~\cite{Jiang2013} or implementation~\cite{Buchholz2022Code}). 
This stays in conflict with \gref{goal_feasibility}.
Other approaches like \gls{geo-ind}\cite{Andres2013}, aimed at location privacy, are not directly applicable to trajectory privacy due to location correlation, as outlined in \gref{goal:uop}.
The predictive mechanism~\cite{Chatzikokolakis2014} addresses these correlations.
The next location is predicted based on the previously protected locations.
Only a small privacy budget is used to test if the predicted location is close to the next real location.
Then, the location is released without using further budget.
Otherwise, a standard noise mechanism is used to perturb the next location, spending more budget.
While the predictive mechanism significantly reduces the required privacy budget from $n\varepsilon$, necessary to protect a trajectory with standard \gls{geo-ind} and \gls{dp} composition, it still causes severe utility degradation for long trajectories (\gref{goal_utility}).
An advantage of these approaches is strong \gls{ldp} guarantees (\refer \secref{goal_guarantees}).
 
\noindent\ballnumber{E} \textbf{Grid-based Approaches}, such as DPT~\cite{He2015}, TGM~\cite{Ghane2020}, and \citet{Sun2023}, first divide the map covered by the dataset into grid cells.
Then, the transition probabilities between cells are recorded, and synthetic trajectories are constructed based on these.
Grid usage allows blocking certain cells, e.g., based on environmental constraints.
However, the accuracy of the generated trajectories depends on the size of the smallest grid cells, and increased granularity yields increased computational effort (\refer \secref{goal_utility}). 
Moreover, capturing movements within grid cells is not possible.
While grid-based approaches offer some appealing properties, they limit the achievable level of utility (\gref{goal_utility}) \cite{TSG}.

\noindent\ballnumber{F}\subheading{Environmental Constraints} 
Existing methods often overlook environmental constraints (\refer \secref{sec_environment}), resulting in trajectories that ignore geographical features like road networks~\cite{Naghizade2020, GeoPointGAN, Miranda-Pascual2023, Sun2023, RAoPT}.
\citet{Haydari2022} address this shortcoming by incorporating geographical constraints through \gls{dp}-based map matching.
Yet, this approach shares the \gls{em} application issue of the aforementioned clustering-based approaches, violating \gref{goal_guarantees}.
The use of dataset-dependent candidate paths for the \gls{em} contradicts the need for dataset-independent outputs (\pitfall{em}).
\citet{Cunningham2021-DPGEN} leverage the publicly available road network to enhance the quality of generated locations. 
However, they produce sets of locations, not trajectories, making trajectory reconstruction from outputs non-trivial.
Despite focusing on locations rather than trajectories, we include this work as an example of successful environmental constraint incorporation for \gls{dp}.

\noindent\ballnumber{G}\subheading{\glsentryshort{poi} protection}
\citet{Cunningham2021-LDP} also accounts for environmental constraints, utilising public knowledge to refine outputs. 
Moreover, this approach offers \gls{ldp} for enhanced guarantees compared to standard \gls{dp}.
However, it is limited to \gls{poi} datasets, making it unsuitable for general coordinate trajectories (\refer \secref{sec_trajectory}).

\subheading{Conclusion}
Despite the development of various trajectory protection mechanisms, no optimal solutions exist.
Many approaches lack sufficient privacy due to incorrect \gls{uop} selection (\gref{goal:uop})~\cite{Jiang2013}, flawed \gls{dp} proofs (\gref{goal_guarantees})~\cite{Hua2015, Li2017, Chen2020}, or susceptibility to reconstruction attacks (\gref{goal_practical})~\cite{Shao2020, RAoPT}.
Additionally, high utility is hard to achieve, with methods releasing only aggregated data~\cite{Hua2015, Li2017, Chen2020, Liu2021a} or relying on grid cells~\cite{He2015, Ghane2020, Sun2023}, restricting granularity.
These limitations inspired researchers to explore \gls{dl}-based trajectory generation as a promising alternative \cite{Liu2018a}, discussed in the following.

\section{DL-based Trajectory Generation}\label{sec_trajGen}

\begin{table*}
    \centering
    \begin{tabular}{l l c c c c c c l}
        \toprule
        \textbf{Category} & \textbf{Approach} & \textbf{\gls{uop}} & \textbf{\ref{goal_guarantees}} & \textbf{\ref{goal:uop}} & \textbf{\ref{goal_practical}} & \textbf{\ref{goal_utility}} & \textbf{\ref{goal_feasibility}} & \textbf{Main Shortcoming} \\
        \midrule
        \multirow{7}{*}{\glsentryshort{lstm}-based} & \ballnumber{1} LSTM-TrajGAN~\cite{Rao2020} & Instance & \xmark & $\checkmark$ & $\checkmark$ (\gls{tul}) / \xmark (\gls{RAoPT}) & $\checkmark$ & $\checkmark$ & No privacy guarantees \\
        & \ballnumber{2} Shin2023~\cite{Shin2023} & Instance & \xmark & $\checkmark$ & $\checkmark$ (\gls{tul}) & $\checkmark$ & $\checkmark$ & Inherited from LSTM-TrajGAN \\
        & \ballnumber{3} Ozeki2023~\cite{Ozeki2023} & Instance & \xmark & $\checkmark$ & $\circ$ (\gls{mia}) & $\checkmark$ & $\checkmark$ & Inherited from LSTM-TrajGAN \\
        & \ballnumber{4} Song2023~\cite{Song2023} & Instance & \xmark & $\checkmark$ & $\checkmark$ (\gls{tul}) & $\checkmark$ & $\checkmark$ & Inherited from LSTM-TrajGAN \\
        & \ballnumber{5} Fontana2023~\cite{Fontana2023} & Instance & \xmark & $\checkmark$ & $\checkmark$ (\gls{tul}) & $\checkmark$ & $\checkmark$ & Inherited from LSTM-TrajGAN \\
        & \ballnumber{6} LGAN-DP~\cite{lgan-dp} & Instance & \xmark & $\checkmark$ & $-$ & $\circ$ & $\checkmark$ & Flawed \gls{dp} proof (\pitfall{domain})\\
        & \ballnumber{7} DP-TrajGAN~\cite{dp-trajgan} & Instance & \xmark & $\checkmark$ & $-$ & $\circ$ & $\checkmark$ & Flawed \gls{dp} proof (\pitfall{domain})\\\hline
        \glsentryshort{aae}-based & \ballnumber{8} Kim2022~\cite{Kim2022} & Location &  & \xmark & $-$ & $\circ$ & $\checkmark$ & \gls{uop}; Grid-based \\ \hline
        Clustering-based & \ballnumber{9} RNN-DP~\cite{Chen2020} & Instance & \xmark & $\checkmark$ & $-$ & $\checkmark$ & $\checkmark$ & Flawed \gls{dp} proof (\pitfall{em})\\ \hline
        \multirow{2}{*}{Two Stage GANs} & \ballnumber{10} \glsentryshort{TSG}~\cite{TSG} & Instance & \xmark & $\checkmark$ &  $-$ & $\checkmark$ & $\checkmark$ & No guarantees\\
        & \ballnumber{11} TS-TrajGEN~\cite{Jiang2023} & Instance & \xmark & $\checkmark$ & $-$  & $\checkmark$ & $\checkmark$ & No guarantees\\ \hline
        Point Generation & \ballnumber{12} GeoPointGAN~\cite{GeoPointGAN} & Location & & \xmark & $-$ & $\checkmark$ & $\checkmark$ & Targets location generation \\
        \bottomrule
    \end{tabular}
    \caption{
    Overview of Trajectory Generation Approaches.
    The table lists the targeted \gls{uop}, adherence to design goals from \secref{sec_goals}, and primary shortcomings.
    Symbols used are $\checkmark$~(goal satisfied), $\circ$~(partially satisfied), and \xmark~(unsatisfied).
    Regarding \textbf{\ref{goal_guarantees}}, \xmark~denotes known \gls{dp} proof errors; the absence of a symbol means we are unaware of any flaw.
    For \textbf{\ref{goal_practical}}, \xmark~indicates demonstrated attacks and approaches with $-$ have not been evaluated against practical attacks.
    }
    \label{tab_trajgan}
    \vspace{-1.5em}
\end{table*}

In 2018, a vision paper~\cite{Liu2018a} proposed the idea of using a \glsfirst{gan} (\refer \secref{sec_gan}) to generate synthetic trajectory data.
This concept relies on the principle that releasing synthetic data safeguards privacy by distributing generated \textit{fake} data, not real individuals' data.
This section explores implementations of the trajGAN concept summarised in \tabref{tab_trajgan}.

\subsubsection*{\ballnumber{1} LSTM-TrajGAN}\label{sec_lstm-trajgan}
The most notable implementation of this concept is \textit{LSTM-TrajGAN~\cite{Rao2020, Rao2020_code}}, which consists of a trajectory generator and a trajectory discriminator with similar architectures.
The generator aims to produce realistic trajectories, while the discriminator seeks to differentiate them from actual trajectories (ref. \secref{sec_gan}).
Real trajectories are normalised by encoding location offsets from a central point and using one-hot encodings for temporal and semantic attributes. 
These encoded trajectories are fed into the generator's embedding layer, which embeds each feature individually.
A \gls{fc} feature fusion layer fuses these embeddings with a noise vector.
A sequence-to-sequence \gls{lstm} layer processes the resulting latent representation. 
Finally, synthetic trajectories are generated using a \gls{fc} layer for continuous properties, like locations, and softmax layers for categorical properties. 
The discriminator, mirroring the generator's structure with embedding, feature fusion, \gls{lstm}, and output layers, evaluates these synthetic trajectories during training, returning a value in the range $[0,1]$ for each input.
LSTM-TrajGAN employs a specialised loss function, \textit{TrajLoss}, as an alternative to the standard adversarial loss, defined by:
\begin{align}\label{eq:trajLoss}
\text{TrajLoss}(y_r, y_p, t_r, t_s) &= \alpha L_{BCE}(y_r, y_p) + \beta L_s(t_r, t_s) \\
&\quad + \gamma L_t(t_r, t_s) + c L_c(t_r, t_s)  \notag
\end{align}
In this equation, $L_{BCE}$ represents the standard adversarial loss, $L_s$ is a \gls{mse} loss measuring spatial similarity, and $L_t$ and $L_c$ are cross-entropy losses assessing temporal and categorical similarities, respectively.

LSTM-TrajGAN demonstrates remarkable utility and is the first to evaluate against \gls{tul} practically, reducing its top-1 accuracy success from \SI{93.8}{\%} to \SI{45.9}{\%}.
Despite its important contribution, this approach faces limitations.
Primarily, it lacks formal privacy guarantees, violating \gref{goal_guarantees}.
A practical evaluation is valuable but insufficient (ref. \secref{goal_guarantees}).
LSTM-TrajGAN is robust against the MARC~\cite{marc2020} \gls{tul} model, but it might be susceptible to other attacks or improved \gls{tul} algorithms.
Secondly, the model's architecture raises further privacy concerns.
The model uses an encoded real trajectory as generation input, with noise only concatenated and intermingled with the remaining input in the \gls{fc} feature fusion layer.
This structure allows the model to learn to disregard the noise to generate more realistic outputs, as detailed in Appendix~\ref{sec_mwe-proof} theoretically and practically in Appendix~\ref{sec_lstm-conv}.
Training LSTM-TrajGAN for the specified $2,000$ batches~\cite{Rao2020} results in differing prediction outputs: about \SI{8}{\%} of predicted hours vary, and the average location distance exceeds \SI{800}{\meter} on the \gls{fs} dataset (\refer \tabref{tab_lstm-conv}).
Extending training to $10\times$ this amount, which only takes $\approx\SI{25}{\min}$ on our system (specified in \secref{sec_generative-models}), leads to $0$ differently predicted hours and reduces the location distance to \SI{183}{\meter}.
This suggests that LSTM-TrajGAN may learn to neglect the input noise with longer training, yielding outputs nearly identical to the input trajectories.
Hence, LSTM-TrajGAN's privacy relies heavily on the model not being trained for too long, a decision dependent on the end-user, and not providing robust guarantees.
It could be argued that LSTM-TrajGAN's latent space acts as a bottleneck, akin to \gls{ae} architectures, preventing precise replication of original trajectories.
However, quantifying the modification caused by this compression is challenging and should not represent the primary basis for the provided privacy preservation.

\subheading{Reconstruction Attack}\label{sec_RAoPTvsLSTM-TrajGAN}
We evaluated the performance of the \gls{RAoPT}~\cite{RAoPT} reconstruction attack on samples produced by LSTM-TrajGAN.
This attack significantly reduces distances between the original and generated trajectories: 
over \SI{20}{\%} for \gls{fs} and between \SI{33}{\%} and \SI{50}{\%} for Geolife. It also improved the overlap of the convex hulls, determined by the Jaccard index, by over \SI{60}{\%} for Geolife and by over \SI{14}{\%} for \gls{fs}.
For detailed measurements and discussion, see Appendix~\ref{sec_raopt-vs-lstm}. 
These findings underscore the insufficiency of practical guarantees alone, as elaborated above.
Although LSTM-TrajGAN effectively protects against the \gls{tul} attack, it remains susceptible to other attacks, like \gls{RAoPT}.

\noindent\textbf{Is LSTM-TrajGAN a GAN?}\label{sec_LSTMonlyAE}
Lastly, we claim that LSTM-TrajGAN operates mainly as a transformative model with minimal use of the discriminator.
To verify this, we removed the discriminator feedback from the loss function ($L_{BCE}$ in Equation~\ref{eq:trajLoss}) and evaluated the model with and without a discriminator on the same dataset (\gls{fs}) used in~\cite{Rao2020}.
\tabref{tab_lstm-loss} in Appendix~\secref{sec_lstm-loss} shows the results.
Performance remains similar whether or not the discriminator's feedback is included in the loss.  
Location quality drops slightly without the discriminator, but categorical properties (e.g., hour) improve.
Suppose we remove all components of the loss and train the generator based only on the discriminator's feedback. In that case, the model provides barely any utility, outputting over \SI{85}{\%} different categorical values (compared to \SI{1}{\%}-\SI{10}{\%} before), and increasing the location distances more than $20$-fold.
In other words, the discriminator has little impact on the model's performance.

\subsubsection*{LSTM-TrajGAN Extensions}\label{sec_ae-models}
Due to the promising results obtained by LSTM-TrajGAN, especially in terms of utility and the openly available source code, several works~\cite{lgan-dp, dp-trajgan, Shin2023, Song2023, Fontana2023} aim to improve the model.
\ballnumber{2}~\textit{\citet{Shin2023}} introduced a \gls{tcac} to counteract mode collapse, a common issue in \glspl{gan}.
However, unlike conventional \glspl{gan} that rely solely on noise input, LSTM-TrajGAN uses real samples as generator inputs, avoiding mode collapse.
We observed no mode collapse during our experiments.
Thus, the effectiveness of the auxiliary classifier remains unclear, and without access to the source code, we could not verify these claims.
Moreover, the approach does not improve on LSTM-TrajGAN's privacy limitations.

\ballnumber{3} The model used by \textit{\citet{Ozeki2023}} appears identical to LSTM-TrajGAN, mirroring even parameters and loss function.
Their model generates privacy-preserving trajectories for taxi demand prediction. 
Yet, without any alterations from LSTM-TrajGAN, the same limitations persist in their approach.
The authors claim their solution would offer equivalent privacy to \gls{dp}.
However, this statement is based on an evaluation regarding the \gls{mia} success rate, which is problematic for two reasons:
First, using one existing attack does not allow extending the findings to any attack, as discussed in \secref{goal_guarantees}.
Second, the approach is compared to CNoise~\cite{Jiang2013} and Geomasking, which do not provide trajectory-level \gls{dp} as discussed in \secref{sec_protection},
Consequently, it is uncertain how their method measures up against an instance-level \gls{dp} mechanism.

\ballnumber{4} \textit{\citet{Song2023}} extend LSTM-TrajGAN by an \gls{exGAN}.
This approach allows sensitive regions to be excluded from the generated data by specifying certain labels to be excluded.
The model is evaluated on the \gls{fs} dataset, where it achieves similar utility to LSTM-TrajGAN while reducing the \gls{tul} success rate.
However, the approach neither adds any privacy guarantees nor addresses the other shortcomings of LSTM-TrajGAN.

\ballnumber{5} \textit{\citet{Fontana2023}} combine LSTM-TrajGAN~\cite{Rao2020} and MARC \cite{marc2020}, a model used for assessing the \gls{tul} success rate by some \gls{lstm}-based approaches~\cite{Rao2020,Shin2023,Song2023}.
Moreover, they introduce a novel \textit{Next Week Trajectory Prediction} model to evaluate the utility of generated trajectories. 
This model mostly equals LSTM-TrajGAN's generator.
However, instead of generating synthetic trajectories, the model receives one week's trajectories and predicts the following week's trajectories for a given user.
This model underlines the adaptability of the LSTM-TrajGAN architecture to other tasks.
Moreover, the authors evaluate the models on two datasets, \gls{fs}, used by most related works, and the breadcrumbs dataset~\cite{breadcrumbs}.
This evaluation confirms the effective balance of utility and privacy of LSTM-TrajGAN.
However, the work does not enhance the generation process and thus shares LSTM-TrajGAN's privacy limitations.

\ballnumber{6} \textit{LGAN-DP}~\cite{lgan-dp} targets the extension of LSTM-TrajGAN by \gls{dp} guarantees.
First, synthetic trajectories are generated through a simplified version of LSTM-TrajGAN, only considering spatial information.
\glsentrylong{dp} is not achieved through the actual generation model but through post-processing.
In particular, a combination of the clustering-based approach by \citet{Hua2015} and the constrained Laplace noise from~\cite{Zhao2019} is implemented.
Both these baselines are discussed in \secref{sec_protection} and have been critiqued for flawed \gls{dp} proofs~\cite{Miranda-Pascual2023}.
Without a comprehensive \gls{dp} proof for LGAN-DP, it likely inherits similar flaws.
As thoroughly discussed by \citet{Miranda-Pascual2023}, \gls{dp} requires adding Laplace noise to all possible outputs, not only to a restricted number of candidates (\pitfall{domain}).
Based on the available information, LGAN-DP appears to construct a candidate set from the cluster centres and only perturbs the corresponding counts, thus failing to meet \gref{goal_guarantees}.
Applying a \gls{dp} protection mechanism to LGAN-DP's outputs likely results in lower utility than directly applying such a mechanism to the real dataset, making the benefits of LGAN-DP over the methods from \secref{sec_protection} unclear.
Moreover, the post-processing is expected to significantly reduce utility compared to other generative approaches such as LSTM-TrajGAN.
We could not confirm this assumption due to unavailable source code.

\ballnumber{7} \textit{DP-TrajGAN~\cite{dp-trajgan}} also targets \gls{dp} guarantees.
Its architecture differs slightly from LSTM-TrajGAN.
However, the general flow remains the same, i.e., a real trajectory concatenated with random noise serves as input for the generator.
Unlike LSTM-TrajGAN, the real trajectory is perturbed to achieve \gls{dp} before being fed into the generator.
This method could ensure \gls{dp} for the output via the post-processing property (\refer \secref{sec_dp}), provided no further steps access raw data.
However, the work states: ''But the discriminator takes the real trajectory as input without adding noise because $D$ needs to identify the original trajectory distribution more accurately, thus providing more accurate weight gradients for $G$'' \cite{dp-trajgan}.
Since the discriminator's feedback indirectly exposes the generator to unprotected trajectories through gradient updates, this contradicts \gls{dp} (\pitfall{domain}).
Thus, DP-TrajGAN fails to offer robust privacy guarantees, violating \gref{goal_guarantees}.

\begin{figure}
    \centering
    \includegraphics[width=\linewidth]{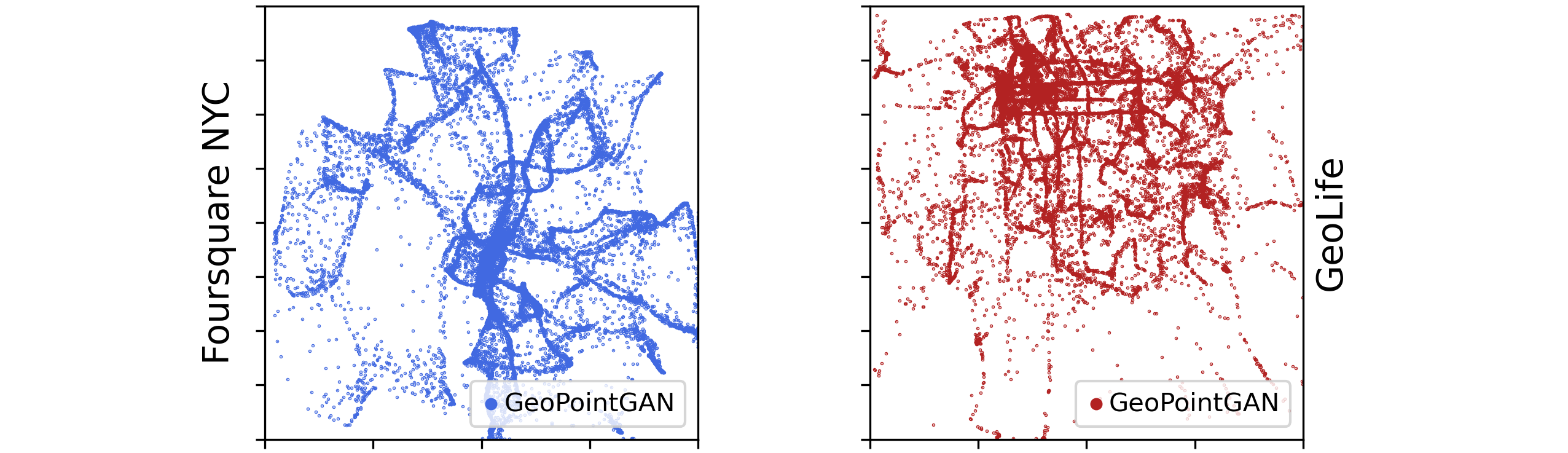}
    \vspace{-0.7cm}
    \caption{
    Application of GeoPointGAN~\cite{GeoPointGAN} to the \gls{fs} and Geolife datasets.
    Compare \figref{fig_gen-pointclouds} for baseline.
    }
    \label{fig_geopointgan}
    \vspace{-1em}
\end{figure}

\subsubsection*{AAE-based Approaches}\label{sec_aae-gen}

While the architectures above mirror LSTM-TrajGAN closely, alternative architectures have been proposed.
\ballnumber{8}~\textit{\citet{Kim2022}} propose a \textit{\gls{aae}-based} architecture for the trajectory generation.
First, the real trajectory data is protected with the established location \gls{dp} mechanism \glsfirst{geo-ind}~\cite{Andres2013}.
Then, the perturbed data is encoded with a \glsfirst{geo-ind}-aware location encoding to offset some of the utility loss.
This client-side protection offers \gls{ldp} guarantees, an advantage over the standard \gls{dp} guarantees of other approaches.
The \gls{aae}'s encoder consists of a many-to-many \gls{lstm} followed by a \gls{fc} layer, while the decoder consists of a \gls{fc} layer followed by an \gls{lstm}.
The discriminator of the \gls{aae} compares the latent space representation produced by the encoder with the real data distribution.
During generation, noise is sampled from a Gaussian distribution and fed into the decoder to generate a synthetic trajectory.
Unlike the \gls{lstm}-based models, this method does not use real trajectories during generation, preventing privacy leaks.
While similar to DP-TrajGAN, the \gls{dp} protection is achieved during pre-processing, the raw data is not accessed in later steps, ensuring \gls{dp} guarantees via the post-processing property (\refer \secref{sec_dp}).

However, the approach has two key drawbacks. 
First, the used privacy mechanism \gls{geo-ind} provides location privacy but not trajectory privacy, violating \gref{goal:uop} (\refer \secref{sec_location-privacy}).
While the usage of an \gls{aae} for trajectory generation adds another layer of privacy on top of \gls{geo-ind}, quantification of the trajectory-level privacy this solution can provide is challenging (conflicting with \gref{goal_guarantees}).
Second, the approach relies on splitting the geographical area into a grid. 
The results provided do not clarify the accuracy of the data generated.
However, the finest used grid, dividing Beijing into a $20 \times 20$ grid, seems coarse.
Grid cell usage limits the provided utility (\refer \gref{goal_utility})~\cite{TSG}.

\subsubsection*{Clustering-based Approaches}\label{sec_cluster-gen}

\ballnumber{9} \textit{\citet{Chen2020}} proposed another \gls{rnn}-based model for trajectory generation.
To provide \gls{dp} guarantees, they post-process the generated trajectories with the trajectory publication mechanism based on clustering proposed by \citet{Hua2015}.
However, this mechanism cannot provide \gls{dp} guarantees due to the incorrect application of the \gls{em} \cite{Miranda-Pascual2023} violating \gref{goal_guarantees} (\pitfall{em}).

\begin{figure*}
    \centering
    \includegraphics[width=\linewidth]{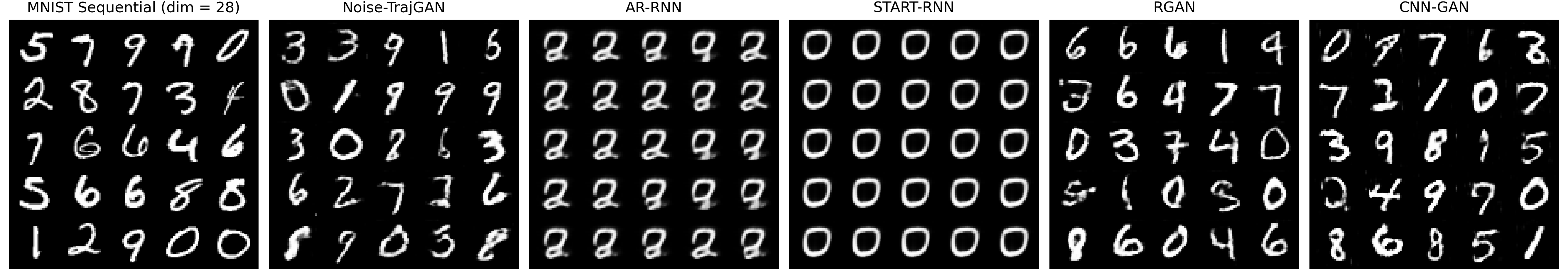}
    \vspace{-0.7cm}
    \caption{
    \gls{mnist-seq} dataset generation.
    All models perform reasonably well on this dataset.
    The two \gls{rnn} models generate the same image as each step depends on the previous, but the first row consists only of black pixels.
    }
    \label{fig_mnist-seq}
    \vspace{-1em}
\end{figure*}

\subsubsection*{Two Stage GANs}\label{sec_grid-gen}

\ballnumber{10} \textit{\citet{TSG}} propose a \gls{TSG} for trajectory generation.
The first stage \gls{gan} consists of a \gls{cnn}, generating a grid representation of a trajectory where each grid cell records the stay duration.
The link module transfers this representation into a sequence of grid cells.
The result serves as input for the second-stage \gls{gan}, which combines it with an encoded road map image to incorporate geographical constraints and outputs coordinate trajectories.
Through the two-stage approach, grid cells can be used to generate the overall structure.
Still, the second stage allows for generating coordinate trajectories for improved accuracy.
While this approach introduces a promising architecture, \gls{TSG} neither provides any privacy guarantees nor is it evaluated against any practical attacks.

\ballnumber{11} \textit{TS-TrajGEN~\cite{Jiang2023}} is a second two-stage \gls{gan} for trajectories.
The first stage generates a region-level (i.e., coarse) trajectory based on recorded origins and destinations.
Based on this regional-level trajectory, the second stage creates a more detailed coordinate Trajectory.
TS-TrajGEN incorporates the road network through the usage of the A*-Search algorithm.
While the approach thoroughly evaluates the utility based on seven metrics, it neither provides privacy guarantees (\gref{goal_guarantees}) nor evaluates the achieved level of privacy (\gref{goal_practical}) in any way.

\subsubsection*{Point Generation}\label{sec_pointgen}

\ballnumber{12} \textit{\glsfirst{gpg}~\cite{GeoPointGAN}} represents a generative model for location sets, not trajectories, which we would like to highlight for its impressive and reproducible point-generation capabilities.
\gls{gpg} employs a classic \gls{gan} structure: the generator uses Gaussian noise to create synthetic points, and the discriminator classifies points as real or synthetic.
For privacy, \gls{gpg} applies \glsfirst{lldp}, a variant of \gls{ldp} treating only labels as sensitive.
Labels indicating whether a location is real or fake are probabilistically flipped to achieve \gls{lldp}.
Our evaluations show that \gls{gpg} accurately captures point distributions of the considered datasets, as depicted in \figref{fig_geopointgan}, and consistently yields stable results.
However, adapting \gls{gpg} for trajectory generation is challenging, as we discuss in \secref{sec_generative-models}.
The model provides location-level privacy by design, so transferring to trajectory-level privacy might significantly degrade utility.
Moreover, while \gls{lldp} yields quantifiable privacy guarantees and some researchers argue that the notion provides sufficient privacy guarantees for the deep learning setting \cite{label_dp, GeoPointGAN}, ultimately, a solution providing standard \gls{dp} guarantees would represent the gold standard.

\subsubsection*{Conclusion}

Several generative models have been published since the first vision paper \cite{Liu2018a} proposing the usage of \glspl{gan} for the privacy-preserving generation of trajectory data. 
However, none of the existing solutions can provide sufficient privacy guarantees yet. 
The largest group of approaches~\cite{Rao2020, Shin2023, Song2023, Fontana2023, Ozeki2023, TSG, Jiang2023} relies on the obfuscation caused by the generation itself.
However, this does not allow quantifying the provided privacy level as discussed in \gref{goal_guarantees}.
Other approaches cannot provide acceptable guarantees due to flaws in their privacy proofs, either due to privacy leakage in the model design \cite{dp-trajgan} or errors inherited from baseline works \cite{Chen2020, lgan-dp}.
The design of a fully differential private generative model remains an open research question.
However, solutions such as \gls{gpg} \cite{GeoPointGAN}, which can provide \gls{lldp} guarantees for robust spatial point generation, show a promising direction.

\section{Generative Models}\label{sec_generative-models}

Current deep learning-based generative models show promise but fail to meet all our design goals (\refer \secref{sec_goals}), with inadequate privacy guarantees as the primary issue (\tabref{tab_trajgan}).
The ideal model would offer
at least \textit{instance-level (preferable user-level)} \textbf{(\ref{goal:uop})}
\textit{differential privacy} guarantees \textbf{(\ref{goal_guarantees})},
while achieving \textit{practical privacy} \textbf{(\ref{goal_practical})}
and \textit{high utility} \textbf{(\ref{goal_utility})}
with \textit{reasonable computational costs}  \textbf{(\ref{goal_feasibility})},
and is evaluated on \textit{public datasets}
with \textit{established metrics}.

\gls{dp-sgd} (\refer \secref{sec_dp-sgd}) is the prevalent method for \gls{dp} in \gls{dl} \cite{dpfyML}, offering instance-level \gls{dp} for training data.
Thus, models trained with \gls{dp-sgd} ensure trajectory-level privacy when each trajectory is a training sample.
The fundamental concept aligns with the objectives of trajectory generation.
\gls{dp-sgd} ensures that models capture overall data distribution while minimising individual samples' impact.
Likewise, trajectory generators aim at encapsulating dataset traits without specific trajectory retention.
This finding motivates the following proposal:
\begin{enumerate*}
    \item Create a generative model independent of input trajectories during prediction, enabling the integration of \gls{dp-sgd}. 
    As \gls{dp-sgd} only protects the training dataset, the method cannot be applied to a model like LSTM-TrajGAN requiring input during prediction.
    \item Upon achieving this, the next step involved applying \gls{dp-sgd} \cite{dpfyML} to ensure \gls{dp} for the dataset.
\end{enumerate*}
In the following, we evaluate six generative models for sequential data to determine their suitability.
All source code is made available\footnote{
    \ifanonymous
        https://github.com/ANONYMIZED
    \else
        \url{https://github.com/erik-buchholz/SoK-TrajGen}
    \fi
}. 

\subheading{Evaluation Setup}\label{sec_hardware}
We performed all measurements on a server (2x Intel Xeon Silver \num{4208}, \SI{128}{\giga\byte} RAM, Ubuntu 20.04.01 LTS) with \num{4} NVIDIA Tesla T4  GPUs (\SI{16}{\giga\byte} RAM each) using one GPU per experiment.
Appendix~\ref{sec_parameters} lists all used (hyper-)parameters.

\begin{figure*}
    \centering
    \includegraphics[width=\linewidth]{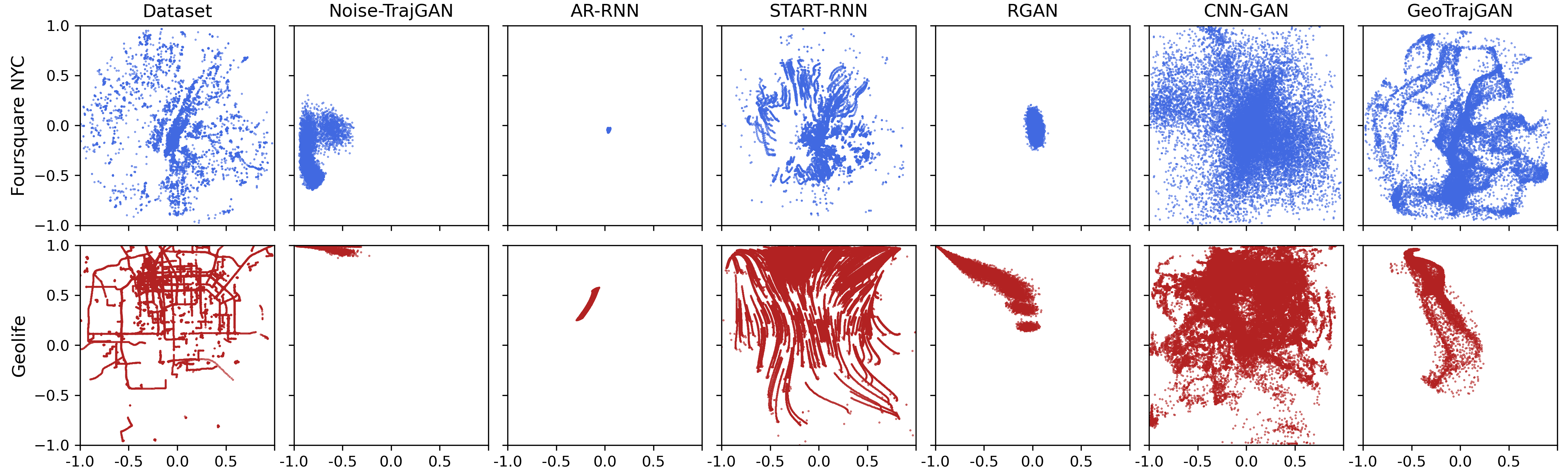}
    \vspace{-0.7cm}
    \caption{
    Overview of generative models. 
    No considered sequential generative model can adequately capture the point distribution of the original trajectory datasets \gls{fs} and Geolife.
    }
    \label{fig_gen-pointclouds}
    \vspace{-1em}
\end{figure*}

\glsreset{mnist-seq}
\subheading{Datasets}\label{sec_eval_datasets}
We use three datasets for our evaluation:
\gls{mnist-seq}~\cite{Esteban2017} as a simple sequential toy dataset to confirm the correctness of our implementations and the two trajectory datasets Geolife~\cite{Geolife1} and \glsfirst{fs}.
While \gls{fs} contains relatively coarse (restaurant) check-ins, Geolife consists of finer trajectories, \SI{91}{\%} having a sampling rate of smaller or equal \SI{5}{\second} \cite{Geolife1}.
The diversity between datasets allows for comparing the approaches for various use cases.
Results on \gls{mnist-seq} are shown in \figref{fig_mnist-seq}, and on trajectory datasets in \figref{fig_gen-pointclouds}.
We detail the three datasets and their pre-processing below.

\subheading{\gls{mnist-seq}}
The standard MNIST dataset consists of handwritten digit images and serves as a benchmark for image processing.
The \gls{mnist-seq} dataset transforms these standard $28\times28$ images into sequences of length $28$ with $28$ values each~\cite{Esteban2017}, i.e., each row of the images represents one time step.
It has been used as a benchmark for sequence generation~\cite{Esteban2017}.
Though not a trajectory dataset, we utilise this simple sequential dataset to confirm our models' capacity to correctly generate sequences, ensuring that weak results on trajectory datasets are not due to implementation errors.

\subheading{FS-NYC}
Like related works \cite{marc2020, Rao2020}, we use \gls{fs} as provided in the LSTM-TrajGAN repository~\cite{Rao2020_code} without further pre-processing.
This version reduces the size of the original dataset~\cite{fs_nyc} to $66\,962$ check-ins and $193$ users and uses categorical features.

\subheading{Geolife}
The Geolife dataset spans a large area of the globe.
Therefore, using it as-is is unsuitable for three reasons:
\begin{enumerate*}[label=\roman*)]
    \item Most existing trajectory generation models target city-scale datasets,
    \item it makes visualisation very challenging, and 
    \item dataset cleaning is required for use with \gls{dl} models.
\end{enumerate*}
We decided on the following pre-processing steps:
\begin{enumerate*}
    \item All points outside a bounding box defined by the 5th ring of Beijing are removed.
    \item All trajectories are re-sampled to a constant sampling rate of one location per \SI{5}{\second} because over \SI{90}{\%} have a finer granularity than that~\cite{Geolife1}.
    \item Trajectories are split on $\geq \SI{60}{\s}$ breaks as we consider separate trips as distinct trajectories.
    \item We truncate all remaining trajectories to $200$ points as some approaches require a defined upper length.
    \item Finally, all trajectories shorter than $10$ points are dropped.
\end{enumerate*}
Both datasets are normalised to $[0;1]$  by subtracting a reference point defined as:
\[
    ref = \left(\frac{\max\{lon\} - \min\{lon\}}{2}, \frac{\max\{lat\} - \min\{lat\}}{2}\right)
\]
and division through a scaling factor defined as:
\[
    sf = \left(\max\{lon\} - ref_{lon}, \max\{lat\} - ref_{lat}\right)
\]

\subheading{\gls{ntg}}\label{sec_ntg}
Considering LSTM-TrajGAN's~\cite{Rao2020} success (\refer \secref{sec_lstm-trajgan}), incorporating \gls{dp-sgd} in its training process may appear promising.
However, LSTM-TrajGAN-based models require real trajectory input for generation, while \gls{dp-sgd} only protects training samples.
Therefore, a mere integration with a \gls{dp-sgd} library is insufficient.
To overcome this limitation, we developed a variant of LSTM-TrajGAN, named \textit{\gls{ntg}}, adhering to a more traditional \gls{gan} structure.
The key difference is that the generator only receives a (Gaussian) noise vector instead of trajectory as input. 
The discriminator remains mostly unmodified.
\figref{fig_mnist-seq} shows that \gls{ntg} can sequentially generate MNIST digits, although the digits are slightly blurrier than those produced by other models.

However, \figref{fig_gen-pointclouds} shows that the model cannot capture the point distribution of trajectory datasets.
Despite over $20$ trials with varied loss functions (like \glsentryshort{wgan-gp} and modified trajLoss), noise shapes, optimizers, learning rates, and \gls{rnn} layers, no notable improvements emerged.
Based on this finding, we examined LSTM-TrajGAN in more detail, described in \secref{sec_LSTMonlyAE}, and found that the discriminator minimally impacts the model's performance. 
Accordingly, solely relying on the discriminator's feedback does not appear promising.
This insight directed us to explore different architectures.

\subheading{Simple RNN Models}
\gls{rnn}-based models have been applied successfully to various domains.
\textit{\gls{lstm}}~\cite{Hochreiter1997} and \textit{\glsentryshort{gru}} \cite{gru2014} are the most widely used \gls{rnn} types.
Indeed, most related works (\refer \secref{sec_trajGen}) incorporate \glspl{rnn} in their models.
We evaluated multiple simple \gls{rnn}-based models, focusing on \textit{\gls{ar}} and \textit{\gls{start}} in the following.

The initial \gls{rnn} cell of the \textit{\gls{ar}} model receives Gaussian noise and produces a latent encoding, which is processed by an output layer into the desired output format, e.g., $(lat, lon)$.
This serves as input to the next cell alongside the previous hidden (and cell) state.
During training, real values replace each cell's output after every step to prevent error propagation.
This results in high-quality images generated during training for \gls{mnist-seq}.
Yet, in the prediction phase, \gls{ar} consistently yields identical outputs (\refer \figref{fig_mnist-seq}).
The initial row of black pixels, shared by all digits, predetermines subsequent rows, yielding near-identical output.
Moreover, \figref{fig_gen-pointclouds} reveals the model's failure to represent point distributions, possibly because initial noise lacks placement information, causing clustering around the dataset's centre of gravity.

Inspired by these findings, we developed \textit{\gls{start}}, using a dataset's real starting point for guidance. 
As this represents a privacy leakage, a final solution would have to generate start points privately, e.g., using \gls{gpg} (\refer \secref{sec_pointgen}).
On the \gls{mnist-seq} dataset, \gls{start} still only produces identical digits because the digits cannot be distinguished based on the first row. 
Trajectory dataset outputs improve, but \figref{fig_gen-pointclouds} shows challenges in capturing details like road networks.
Consequently, we consider more sophisticated generative models that are successful in other domains.

\subheading{RGAN~\cite{Esteban2017}}
The \textit{\glsfirst{rgan}} \cite{Esteban2017} represented the first application of the \gls{gan} architecture to real-valued sequences, avoiding discretization drawbacks (\refer \secref{sec_grid-utility}).
This model employs a standard \gls{gan} setup:
A generator creates sequences from noise, and a discriminator differentiates these from real training samples.
Both models consist of a \gls{lstm} layer followed by a \gls{fc} output layer.
The input shape is $(batch\_size, sequence\_len, noise\_dim)$, with $noise\_dim$ set to $5$ in \cite{Esteban2017}.
We reproduced the results on \gls{mnist-seq} but observed mode collapse with the original hyper-parameters, yielding only some of the $10$ digits.
Applying \glsentryshort{wgan-lp}~\cite{Petzka2018} improved results in \figref{fig_mnist-seq}.
Nevertheless, \gls{rgan} failed on the trajectory datasets (\refer \figref{fig_gen-pointclouds}).
Though effective for sequential data like \gls{mnist-seq} and medical signals, \gls{rgan} fails to capture spatial point distributions.
The generated point distributions vary greatly between gradient updates, indicating unstable training.
Multiple practitioners report difficulties in training \glspl{gan} containing \gls{rnn} layers and report better results using \glsentryshort{conv1d} layers instead \cite{rnn_in_gan}.

\subheading{\gls{cnn}-\gls{gan}}
One notable example is WaveGAN~\cite{Donahue2019}, which proposes a \glsentryshort{conv1d}-based \gls{gan} for the generation of audio signals and performs better than compared \gls{rnn}-based models.
WaveGAN represents an adaption of the popular and robust~\cite{ganhacks} DCGAN~\cite{DCGAN} architecture to the 1D setting.
In particular, the $5\times5$ convolutional filters are flattened to \glsentryshort{conv1d} filters with kernel size $25$ and both stride and upsampling are increased from $2$ to $4$~\cite{Donahue2019}.
As WaveGAN produces a minimum sequence length of $16\,384$, we modified the model to accommodate shorter sequences required by \gls{mnist-seq} (length $28$) and our trajectory datasets (lengths up to $200$).
We retained a kernel size of $25$ to enable the model to consider multiple past and future points during generation.
To reduce upsampling, we reduced the stride to $1$.
We calculate the initial \gls{fc} layer's output size based on the dataset's maximum sequence length.
The resulting model, trained with the \glsentryshort{wgan-lp} loss, achieves results comparable to \gls{rgan} on \gls{mnist-seq}.
Although the generated points' distribution is much closer to the dataset's and the model trains more stable than previous models, the \gls{cnn}-\gls{gan} remains unable to replicate the datasets' detailed spatial distributions (\refer \figref{fig_gen-pointclouds}).
However, the improved results indicate that \glsentryshort{conv1d} layers are more suitable for capturing point distributions than \glspl{rnn}.

\subheading{GeoTrajGAN}
A prevailing pattern of the previous experiments is the models' inability to capture the datasets' location distribution.
Therefore, we considered \gls{gpg}~\cite{GeoPointGAN}, discussed in \secref{sec_pointgen}, which can generate detailed point distributions (\refer \figref{fig_geopointgan}) and intended to extend it to a trajectory generation model called \textit{\gls{gtg}}.
The first challenge relates to the selection of batches.
Originally, \gls{gpg} randomly samples $7\,500$ points per batch, so each batch resembles the dataset's distribution.
We found that using randomly sampled trajectories and concatenating their points already impairs the model's ability to represent point distributions accurately.
We attributed this to the use of batch norm, as the trajectory batches provide less diversity than point batches because locations within one trajectory are correlated.
Replacing batch norm by layer norm solved this problem but required reducing the learning rate from 4E-5 to 1E-6 and adding a \textit{\glsentryshort{stn}} to the discriminator.
In contrast, \gls{gpg} only uses an \glsentryshort{stn} in the generator.
These changes increase the discriminator's parameter count $12.9\times$, slowing down training.
\gls{gtg} takes $\approx500$ epochs to reach a comparable state to \gls{gpg} after $100$ epochs.
Moreover, the street network details are not as sharp when using layer norm, but most roads remain visible.
However, the resulting model still produced points only.
To produce trajectories, we tried several modifications, achieving the best results with a bidirectional \gls{lstm} in the generator and two separate discriminators: 
one similar to \gls{gpg}'s (with layer norm adjustments) and a new \gls{lstm}-based discriminator for assessing the sequential quality.

\figref{fig_gen-pointclouds} shows that \gls{gtg} somewhat captures \gls{fs}'s point distribution but still fails on Geolife despite its larger size and longer training.
Moreover, the generated distributions do not reach the detail captured by the original \gls{gpg} model (\refer \figref{fig_geopointgan}).
While being the most promising of all evaluated models, the \gls{gtg} approach requires substantial future work to be suitable for privacy-preserving trajectory generation.
Nevertheless, the findings highlight that an ensemble model consisting of a model specialised in capturing the distribution of locations and another model capturing sequential properties appears to be a promising direction for future research.

\subsubsection*{Discussion}
Our evaluations highlight that applying sequential \gls{gan} models from other domains to trajectory datasets is challenging.
We identified the application of \gls{dp-sgd} as a promising methodology for generating trajectories with rigorous privacy guarantees.
Relying on this well-researched algorithm could solve the privacy limitations of existing generative trajectory models (\refer \secref{sec_trajGen}).
However, to enable the application of \gls{dp-sgd}, developing a generative model that does not access the real dataset during generation is essential. 
We are unaware of an existing model achieving this for trajectories with a continuous location domain.
Moreover, we empirically evaluated multiple generative models targeting continuous sequential data but found none could capture the distribution of locations in a trajectory dataset.
The design of an ensemble model combining one part for point distribution and another for sequential properties presents a promising direction for future research.
Additionally, our experiments indicated that \glsentryshort{conv1d} layers perform better than \gls{rnn}-based models for generating trajectories.

Our experiments' main takeaway is that developing a generative model for trajectory datasets with strong semantic privacy guarantees remains an important open research question.
Concrete next steps for future work could be:
\begin{enumerate*}
    \item \label{fr:1} Developing a generative model for trajectory data without real data access during generation. Such a model could be based on a model ensemble favouring \glsentryshort{conv1d} over \gls{rnn} layers.
    \item \label{fr:2} Integrating this model with the \gls{dp-sgd} algorithm to provide trajectory-level privacy guarantees. As \gls{dp-sgd} introduces noise to the training process, further model adaptions might be required.
    \item \label{fr:3} Considering alternative approaches like label \gls{dp}, as used by \gls{gpg}, if the impact of \gls{dp-sgd} on utility is too detrimental.
    \item \label{fr:4} Special purpose solutions for \gls{poi} and discrete trajectories could offer improved utility under full privacy guarantees for certain use cases.
\end{enumerate*}
These developments should be guided by the design framework proposed in \secref{sec_goals}.
With these suggestions for future work, we conclude in the following section.

\section{Conclusion}\label{sec_conclusion}

This study examined the current state-of-the-art privacy-preserving trajectory dataset generation, focusing on deep learning techniques.
We proposed a framework comprised of five design goals to guide the future development of generative trajectory approaches.
We emphasised the ''\glsentrylong{uop}'', which is frequently overlooked in current methods, and proposed a novel systematisation of utility metrics.
Our analysis of current trajectory protection methods revealed a lack of robust solutions, with multiple works grounded in flawed differential privacy proofs.
The influence of these flawed proofs on subsequent research highlights the importance of diligent assessment of such guarantees.
Deep learning-driven generative models reveal their potential as an alternative to protection mechanisms, especially regarding utility.
Yet, they fall short of delivering robust trajectory-level differential privacy guarantees.
Furthermore, our extensive experimental study suggests that generative models designed for other sequential domains are not readily transferable to trajectory datasets.
The main outcome of this work is that the design of a generative deep-learning model providing \gls{dp} guarantees represents a compelling open research question.
We identify a \gls{gan}-based model trained with \gls{dp-sgd} as a promising research direction that we intend to pursue as future work.
\begin{acks}
    The authors would like to thank UNSW, the Commonwealth of Australia,
    and the Cybersecurity Cooperative Research
    Centre Limited for their support of this work.
    The authors thank all the anonymous reviewers for their valuable feedback.
\end{acks}

\bibliographystyle{ACM-Reference-Format}
\bibliography{library}

\appendix

\begin{table*}
    \centering
    \begin{tabular}{lrrrrrrr}
        \toprule
                \textbf{Dataset} &  \textbf{Epochs} &  \textbf{Batches} &  \textbf{Haversine Mean [m]} &  \textbf{Euclidean Mean} &  \textbf{Hour [\%]} &  \textbf{Day [\%]} &  \textbf{Category [\%]} \\
        \midrule
         Foursquare NYC &     250 &     2000 &              837.74 &            1.50 &      7.78 &     0.01 &          0.79 \\
         Foursquare NYC &    2500 &    20000 &              182.99 &            0.35 &      0.00 &     0.00 &          0.01 \\
                GeoLife &      11 &     2000 &              492.06 &            8.52 &      3.24 &     0.01 &           N/A \\
                GeoLife &     110 &    20000 &              132.58 &            2.51 &      0.01 &     0.00 &           N/A \\
        GeoLife Spatial &      11 &     2000 &              230.02 &            4.12 &       N/A &      N/A &           N/A \\
        GeoLife Spatial &     110 &    20000 &               92.28 &            1.57 &       N/A &      N/A &           N/A \\
        \bottomrule
        \end{tabular}
    \caption{LSTM-TrajGAN~\cite{Rao2020} Convergence. 
    The generated trajectories converge towards the generation inputs if the model is trained long enough.
    The table records the mean distances of the predicted values from the original dataset after $2\,000$ batches as used in the paper~\cite{Rao2020} and after training $10\times$ as long, i.e., after training for $20\,000$ batches.
    }
    \label{tab_lstm-conv}
\end{table*}

\glsadd{adamw}
\printglossaries

\begin{table*}
\centering
\begin{tabular}{llrrrrr}
\toprule
        Dataset &     Loss &  Haversine Mean [m] &  Euclidean Mean &  Hour [\%] &  Day [\%] &  Category [\%] \\
\midrule
 Foursquare NYC &      STD &              762.65 &            1.35 &      9.05 &     0.02 &          0.78 \\
 Foursquare NYC &   NO BCE &              788.88 &            1.36 &      5.56 &     0.00 &          1.42 \\
 Foursquare NYC & BCE ONLY &            14973.26 &           26.09 &     95.71 &    86.82 &         90.90 \\
        GeoLife &      STD &              451.00 &            7.82 &      2.49 &     0.01 &           N/A \\
        GeoLife &   NO BCE &              440.32 &            7.54 &      2.70 &     0.01 &           N/A \\
        GeoLife & BCE ONLY &             9090.44 &          163.93 &     94.13 &    85.80 &           N/A \\
GeoLife Spatial &      STD &              191.21 &            3.64 &       N/A &      N/A &           N/A \\
GeoLife Spatial &   NO BCE &              276.48 &            4.73 &       N/A &      N/A &           N/A \\
GeoLife Spatial & BCE ONLY &             6403.59 &          126.27 &       N/A &      N/A &           N/A \\
\bottomrule
\end{tabular}
\caption{Evaluation of LSTM-TrajGAN~\cite{Rao2020} with different losses.
The results show that the discriminator has a minimal impact on model performance and that the \glsentryshort{bce}-loss alone does not yield reasonable utility.
}
\label{tab_lstm-loss}
\end{table*}

\section{Minimal Example trajLoss}\label{sec_mwe-proof}

In \secref{sec_trajGen}, we describe LSTM-TrajGAN's~\cite{Rao2020} potential to ignore the input noise. 
This is due to noise being added by concatenation, followed by a \gls{fc} layer. 
We demonstrate this with a simplified example:
\begin{enumerate}
    \item $x$ is the encoded input of shape $batch\_size \times len \times dim$
    \item $n$ is the random noise of shape $batch\_size \times noise\_dim$
    \item Generator $G(x, n)$: $G: \mathbb{R}^{len \times (dim + noise\_dim)}\to \mathbb{R}^{len \times dim}$
    \item Discriminator $D(x)$: $D: \mathbb{R}^{len \times dim} \to [0, 1]$
\end{enumerate}
The noise vector $n$ is repeated for all time steps $len$ to map the shape $batch\_size \times noise\_dim$ to $batch\_size \times len \times nosie\_dim$.
LSTM-TrajGAN uses \textit{TrajLoss} as loss function~\cite{Rao2020}:
\begin{align}
\text{TrajLoss}(y_r, y_p, t_r, t_s) &= \alpha L_{BCE}(y_r, y_p) + \beta L_s(t_r, t_s) \\
&\quad + \gamma L_t(t_r, t_s) + c L_c(t_r, t_s) \notag
\end{align}
For generator training, $y_r$ is $1$, $y_p=D(G(x,n))$, $t_r$ is the real sample ($t_r = x$), and $t_s$ is the generated sample ($t_s = G(x, n)$).
The generator's goal is to minimise this loss.
\begin{align}
    & \argmin_{G(x,n)} &\text{TrajLoss}(1, D(G(x,n)), x, G(x,n)) \\
    = & \argmin_{G(x,n)} &\alpha L_{BCE}(1, D(G(x,n)) + \beta L_s(x, G(x,n)) + \\ \label{eq:min-trajloss}
    && \gamma L_t(x, G(x,n)) + c L_c((x, G(x,n)) \notag
\end{align}
The spatial loss $L_s$ is the \gls{mse}, computed as $MSE(x, G(x,n))$ for generator training.
This part of the loss is minimised for $G(x, n) = x$, as $MSE(x,x)=0$.
Temporal similarity $L_t$ and $L_c$ are softmax cross entropies, also minimised for $G(x, n) = x$.
With an optimally trained discriminator, the adversarial loss $L_{BCE}$ is minimised for $G(x, n) = x$, as in this case, the real and generated samples are identical.
This leads in:

\begin{equation}
\argmin_{G(x,n)}  [ \text{TrajLoss}(1, D(G(x,n)), x, G(x,n)) ]  =  x
\end{equation}
The generator may learn to ignore the noise with sufficient training.
The model employs a \gls{fc} fusion layer $F$, merging encoding $x$ with noise $n$: $F(x,n) = W (x||n) + b$. 
This formula simplifies to $F(x,n) = W_x x + W_n n + b$.
As $x$ and $n$ are only concatenated, we can simplify this to $F(x,n) = W_x x + W_n n + b$.
With prolonged training, if the loss is minimal for $F(x,n) = x$, weights in $W_n$ related to $n$ will tend towards zero. 
Although LSTM-TrajGAN's generator $G$ is more complex than $F$, the key fusion component in $G$ operates similarly to $F$.
Given that the fusion layer is biased towards ignoring the noise, we infer this issue could extend to LSTM-TrajGAN's more intricate structure. 
This is concerning for privacy, as it suggests that the model's noise input does not ensure the output trajectories differ significantly from the input trajectories. 
Especially with a large latent space, the model could end up reproducing the real trajectories it was fed instead of creating distinct synthetic ones. 
We demonstrate this issue practically with experiments in Appendix~\ref{sec_lstm-conv}.

\section{LSTM-TrajGAN Evaluations}\label{apendix:lstm-evals}

In \secref{sec_lstm-trajgan}, we report on three evaluations performed with LSTM-TrajGAN~\cite{Rao2020}.
In this section, we describe these measurements in more detail.
All experiments described in this section have been recorded with $5$-fold cross-validation on the system specified in \secref{sec_hardware}.
Except for those parameters specified in the following sections, we used the parameters as specified in the respective publications.
The \gls{fs} and Geolife datasets introduced in \secref{sec_datasets} are used for all measurements.
The dataset denoted as \textit{Geolife Spatial} is equivalent to the Geolife dataset except that only the spatial information, i.e., latitude and longitude, are used, while temporal information is ignored.
\secref{sec_lstm-conv} describes the evaluation testing whether LSTM-TrajGAN can learn to ignore the input noise. 
In \secref{sec_lstm-loss}, we evaluate the influence of the discriminator's feedback on LSTM-TrajGAN's training.
Finally, \secref{sec_raopt-vs-lstm} evaluates the effectiveness of \gls{RAoPT}, proposed in~\cite{RAoPT}, on LSTM-TrajGAN.

\subsection{LSTM-TrajGAN Convergence}\label{sec_lstm-conv}

In \secref{sec_lstm-trajgan}, we describe that the LSTM-TrajGAN's architecture does not guarantee that the input noise influences the generated trajectories.
This is because the input noise is only concatenated to the input trajectory before being processed by \gls{fc} feature fusion layer, allowing the model to learn to ignore the noise.
We elaborate on this theoretically in \secref{sec_mwe-proof}.
To show this behaviour practically, we performed the following evaluation.
First, we trained LSTM-TrajGAN for $2\,000$ batches, as done in the original paper~\cite{Rao2020}, on the \gls{fs} and the Geolife datasets and recorded the difference between the input trajectories of the test set and the generated trajectories. 
Concretely, we measured the haversine and Euclidean distances between each pair of locations and the percentage of all categorical values that differ between input and output trajectories.
Then, we trained the mode for $10\times$ as long, i.e., $20\,000$ batches and recorded the same values.
The results for these measurements are provided in \tabref{tab_lstm-conv}.
If LSTM-TrajGAN is trained for $2\,000$ batches, the results show a clear difference between the generation inputs and the generated trajectories. 
On the \gls{fs} dataset, distances differ by over \SI{800}{\meter} on average, and $\approx$\SI{8}{\%} of hour values are different in the output.
On the Geolife dataset, the predicted locations have an average distance of $\approx$\SI{500}{\meter} if the time is part of the model and \SI{230}{\meter} if only the spatial information is considered.
After $20\,000$ batches, the differences between input and output trajectories are significantly reduced.
Instead of an average distance of \SI{800}{\meter}, the output locations only differ by \SI{180}{\meter}, which might be within line of sight for many places. 
Moreover, nearly all categorical features are identical to the values from the original dataset. 
The same is true for the Geolife dataset, where the locations are even closer to those from the original dataset, and the categorical features are nearly identical, too.
These results highlight that LSTM-TrajGAN provides little privacy if trained for too long, as the generated trajectories closely resemble the trajectories from the original dataset used as generation inputs.
Thus, as discussed in \secref{sec_lstm-trajgan}, the privacy provided by the model depends on the number of epochs the model is trained for, which provides limited privacy guarantees.

\begin{table*}
    \centering
    \begin{tabular}{lccc}
        \toprule
                Dataset &  Euclidean Improvement [\%] &  Hausdorff Improvement [\%] &  Jaccard Improvement [\%] \\
        \midrule
        GeoLife Spatial &                      33.40 &                      50.15 &                    60.51 \\
                GeoLife &                      34.95 &                      45.57 &                    69.15 \\
         Foursquare NYC &                      24.94 &                      19.81 &                    14.30 \\
        \bottomrule
    \end{tabular}
    \caption{
    The table shows the percentage reduction of the Euclidean and Hausdorff distances and the increase of the Jaccard Index of the convex hull after applying RAoPT~\cite{RAoPT} on a dataset generated with LSTM-TrajGAN~\cite{Rao2020}.
    }
    \label{tab_raopt-vs-lstm}
    \vspace{-2em}
\end{table*}

\subsection{LSTM-TrajGAN Loss}\label{sec_lstm-loss}

In \secref{sec_LSTMonlyAE}, we raise the question to what extent the model operates as a \gls{gan}~\cite{Goodfellow2014}, and to which extent the modification caused by the model are simply transformative due to the latent space compression of the input.
To evaluate this question, we trained the model with three different loss functions and recorded all results in \tabref{tab_lstm-loss}. 
First, we used the TrajLoss loss (\refer Equation~\ref{eq:trajLoss}) proposed by the authors~\cite{Rao2020} and denoted as \textit{standard (STD)} in \tabref{tab_lstm-loss}.
Second, we excluded the discriminator from the model's training process by setting $\alpha = 0$ in Equation~\ref{eq:trajLoss}, i.e., excluding the \gls{bce}-loss from the loss computation.
This case is denoted as \textit{NO BCE}.
Third, we trained the model with the \gls{bce}-loss only by setting all other factors in TrajLoss to $0$, denoted as \textit{BCE ONLY}.
Again, we record the differences between the input and output trajectories of the test set by measuring the locations' mean haversine and Euclidean distance and the percentage of categorical attributes that differ. 
In this evaluation, lower values are better, as the model tries to generate trajectories that are as close as possible to those used as input (\refer \secref{sec_lstm-conv}).

\tabref{tab_lstm-loss} shows that the difference between the models trained with (\textit{STD}) and without (\textit{NO BCE}) the discriminator is very small.
While on the \gls{fs} and the Geolife Spatial datasets, the location differences are slightly smaller when the model is trained with the discriminator, the model performs even better without the discriminator on the Geolife dataset with categorical features.
Moreover, the results show that the discriminator's feedback alone is insufficient for successfully training the model. 
The mean distance is increased $\approx20\times$ on all datasets when the model is trained with the \gls{bce}-loss only compared to the standard TrajLoss.
Likewise, the difference in categorical features increases from \SI{9}{\%} (hour) and $\leq\SI{1}{\%}$ (day and category) to over \SI{85}{\%} in all cases.
These results highlight that LSTM-TrajGAN trains differently than traditional \glspl{gan}, such as DCGAN~\cite{DCGAN}, and explain why our \glsentrylong{ntg} (\refer \secref{sec_ntg}) performed so poorly on the trajectory datasets.

\subsection{RAoPT on LSTM-TrajGAN}\label{sec_raopt-vs-lstm}

Finally, in \secref{sec_RAoPTvsLSTM-TrajGAN}, we ask whether LSTM-TrajGAN is susceptible to existing attacks against trajectory protection approaches, such as \gls{RAoPT}~\cite{RAoPT}.
To evaluate this, we used the authors' \gls{RAoPT} implementation~\cite{Buchholz2022Code} and ran it on trajectories generated by LSTM-TrajGAN.
Concretely, we performed the measurements as follows.
For each of the $5$ runs, we randomly split the dataset into two equal-sized sets $train_{trajGAN}$ and $test_{trajGAN}$. 
Then, we trained LSTM-TrajGAN for $2\,000$ batches, as used in \cite{Rao2020}, on $train_{trajGAN}$.
After completing the training, we generated trajectories based on $test_{trajGAN}$.
The resulting set of generated trajectories, $gen_{trajGAN}$, was split in proportion $2:1$ into the larger set $train_{RAoPT}$, and the smaller $test_{RAoPT}$. 
While $train_{RAoPT}$ was used for training of the \gls{RAoPT} model, we performed the final measurements on $test_{RAoPT}$ and show the results in \tabref{tab_raopt-vs-lstm}.
As we used the \gls{RAoPT} implementation without modification, we recorded the same values as used in \cite{RAoPT}. 
\gls{RAoPT} can reduce both the Euclidean and the Hausdorff distances and increase the Jaccard index of the trajectories' convex hull, which serves as a representation of the overlap of two trajectories' activity spaces~\cite{RAoPT}. 
While the reconstruction is not as significant as for those protection mechanisms considered in~\cite{RAoPT}, the attack still reduces the Euclidean distance by over \SI{30}{\%} on both variants of the Geolife dataset, and the Hausdorff distance even by over \SI{45}{\%}. 
Moreover, on the Geolife dataset, the Jaccard index can be increased by over \SI{60}{\%} through the reconstruction.
In summary, while LSTM-TrajGAN seems significantly less susceptible to the attack than some traditional trajectory protection mechanisms, the reconstruction still succeeds to some extent.

\section{Model Parameters}\label{sec_parameters}

In this section, we record the (hyper-)parameters used for the evaluation of generative models in \secref{sec_generative-models}, which produce Figures~\ref{fig_mnist-seq} and ~\ref{fig_gen-pointclouds}. 
Note that these parameters are also recorded in the provided code repository\footnote{
    \ifanonymous
        https://github.com/ANONYMIZED
    \else
        \url{https://github.com/erik-buchholz/SoK-TrajGen}
    \fi
}.
We follow the same order as \secref{sec_generative-models}.
An overview of all (hyper-)parameters is provided in \tabref{tab-hyperparameters}.

\subsection{Noise-TrajGAN}

On all datasets, we use a batch size of $10$, PyTorch's \textit{\gls{adamw}} optimiser with a learning rate of $0.0003$ for the discriminator and $0.0001$ for the generator, $\beta_1 = 0.5$ and $\beta_2=0.999$.
We update the discriminator $5\times$ more frequently than the generator.
Both the dimensionality of the input noise vector and the \gls{lstm}'s hidden size are chosen as $100$.
The discriminator's embedding dimensions are chosen as $28$ for \gls{mnist-seq}, $64$ for the combined spatial features, and equal to the input size for categorical features.
We train the models with the \gls{wgan-lp} loss.
On \gls{mnist-seq} and Geolife, we train for $100$ epochs, and on \gls{fs} for $300$ due to its smaller size.
All other (hyper-)parameters are kept as their respective default values.

\subsection{AR-RNN and START-RNN}

Both \gls{rnn}-based models use a batch size of $512$, the \textit{\gls{adamw}} optimiser with a learning rate of $0.001$, $\beta_1 = 0.9$, and $\beta_2=0.999$ for all datasets.
On \gls{mnist-seq}, the models are trained for $30$ epochs, on \gls{fs} for $300$, and on Geolife for $100$.
The dimensionality of the input noise used by \gls{ar} is chosen as the feature dimensionality of the dataset, i.e., $28$ for \gls{mnist-seq} and $2$ for the trajectory datasets.
The initial \gls{fc} layer uses the output size as the input size. 
An \gls{lstm} with a hidden size of $100$ is used for both models.
Both models use the \gls{mse} as the loss function.
All other (hyper-)parameters are kept as their respective default values.

\begin{table*}[ht]
\centering
\begin{tabular}{@{}lllllllllcl@{}}
\toprule
\textbf{Model}       & \textbf{Dataset}         & \textbf{Epochs} & \textbf{Batch Size} & \textbf{Optimizer} & \textbf{LR Gen} & \textbf{LR Dis} & \textbf{$\beta_1$} & \textbf{$\beta_2$} & \textbf{Loss} & \textbf{$n_{critic}$} \\ \midrule
\textbf{\gls{ntg}}   & \textbf{\gls{mnist-seq}} & $100$           & $10$                & \gls{adamw}              & 1E-4            & 1E-4            & $0.5$              & $0.999$            & \glsentryshort{wgan-lp} & $5$                   \\
\textbf{\gls{ntg}}   & \textbf{\gls{fs}}        & $300$           & $10$                & \gls{adamw}              & 1E-4            & 3E-4            & $0.5$              & $0.999$            & \glsentryshort{wgan-lp} & $5$                   \\
\textbf{\gls{ntg}}   & \textbf{Geolife}         & $100$           & $10$                & \gls{adamw}              & 1E-4            & 3E-4            & $0.5$              & $0.999$            & \glsentryshort{wgan-lp} & $5$                   \\
\textbf{\gls{ar}}    & \textbf{\gls{mnist-seq}} & $30$            & $512$               & \gls{adamw}              & 1E-3            & N/A             & $0.9$              & $0.999$            & \gls{mse}     & $1$                   \\
\textbf{\gls{ar}}    & \textbf{\gls{fs}}        & $300$           & $512$               & \gls{adamw}              & 1E-3            & N/A             & $0.9$              & $0.999$            & \gls{mse}     & $1$                   \\
\textbf{\gls{ar}}    & \textbf{Geolife}         & $100$           & $512$               & \gls{adamw}              & 1E-3            & N/A             & $0.9$              & $0.999$            & \gls{mse}     & $1$                   \\
\textbf{\gls{start}} & \textbf{\gls{mnist-seq}} & $30$            & $512$               & \gls{adamw}              & 1E-3            & N/A             & $0.9$              & $0.999$            & \gls{mse}     & $1$                   \\
\textbf{\gls{start}} & \textbf{\gls{fs}}        & $300$           & $512$               & \gls{adamw}              & 1E-3            & N/A             & $0.9$              & $0.999$            & \gls{mse}     & $1$                   \\
\textbf{\gls{start}} & \textbf{Geolife}         & $100$           & $512$               & \gls{adamw}              & 1E-3            & N/A             & $0.9$              & $0.999$            & \gls{mse}     & $1$                   \\
\textbf{\gls{rgan}}  & \textbf{\gls{mnist-seq}} & $300$           & $28$                & \gls{adamw}              & 1E-4            & 1E-4            & $0.5$              & $0.999$            & \glsentryshort{wgan-lp} & $5$                   \\
\textbf{\gls{rgan}}  & \textbf{\gls{fs}}        & $500$           & $28$                & \gls{adamw}              & 1E-4            & 1E-4            & $0.5$              & $0.999$            & \glsentryshort{wgan-lp} & $5$                   \\
\textbf{\gls{rgan}}  & \textbf{Geolife}         & $100$           & $28$                & \gls{adamw}              & 1E-4            & 1E-4            & $0.5$              & $0.999$            & \glsentryshort{wgan-lp} & $5$                   \\
\textbf{CNN-GAN}     & \textbf{\gls{mnist-seq}} & $300$           & $32$                & \gls{adamw}              & 1E-4            & 3E-4            & $0.5$              & $0.999$            & \glsentryshort{wgan-lp} & $1$                   \\
\textbf{CNN-GAN}     & \textbf{\gls{fs}}        & $1\,000$         & $32$                & \gls{adamw}              & 1E-4            & 3E-4            & $0.5$              & $0.999$            & \glsentryshort{wgan-lp} & $1$                   \\
\textbf{CNN-GAN}     & \textbf{Geolife}         & $300$           & $32$                & \gls{adamw}              & 1E-4            & 3E-4            & $0.5$              & $0.999$            & \glsentryshort{wgan-lp} & $1$                   \\
\textbf{\gls{gtg}}   & \textbf{\gls{fs}}        & $1\,000$         & $256$               & \gls{adamw}              & 1E-5            & 1E-5            & $0.5$              & $0.999$            & \gls{bce}       & $1$                   \\
\textbf{\gls{gtg}}   & \textbf{Geolife}         & $1\,000$         & $256$               & \gls{adamw}              & 1E-5            & 1E-5            & $0.5$              & $0.999$            & \gls{bce}       & $1$                   \\ \bottomrule
\end{tabular}
\caption{
Evaluation Hyperparameters.
The table shows the hyperparameters of the models from \secref{sec_generative-models} used for the evaluation and Figures~\ref{fig_mnist-seq} and~\ref{fig_gen-pointclouds}.}
\label{tab-hyperparameters}
\end{table*}

\subsection{RGAN}

The \gls{rgan} uses a batch size of $28$, and PyTorch's \textit{\gls{adamw}} optimiser with $\beta_1=0.5$ and $\beta_2=0.999$ for both generator and discriminator. 
On \gls{mnist-seq} and Geolife, both learning rates are chosen as $0.0001$, while the discriminator is trained with a learning rate of $0.0003$ on \gls{fs}. 
All cases use the \gls{wgan-lp} loss and train the discriminator $5\times$ per generator update.
For the input noise, a dimension of $5$ is used as in the original paper~\cite{Esteban2017}, and the \gls{lstm}'s hidden size is chosen as $100$.
On \gls{mnist-seq}, the model is trained for $300$ epochs, on \gls{fs} for $500$, and on Geolife for $100$.
All other (hyper-)parameters are kept as their respective default values.

\subsection{CNN-GAN}

Our \gls{cnn}-gan uses a batch size of $32$ on all datasets, PyTorch's \textit{\gls{adamw}} optimiser with $\beta_1=0.5$ and $\beta_2=0.999$ and a learning rate of $0.0003$ for the discriminator and $0.0001$ for the generator.
The generator and discriminator are updated with the same frequency ($n_{critic}=1$), and \gls{wgan-lp} is used as the loss.
On \gls{mnist-seq} and Geolife, the model is trained for $300$ epochs, and on \gls{fs} for $1\,000$ epochs.
For the input noise, a shape of $(batch\_size, 100)$ is used, i.e., the noise has a dimensionality of $100$.
The model uses batch norm on \gls{mnist-seq}, while no batch norm is used on the trajectory datasets.
All other (hyper-)parameters are kept as their respective default values.

\subsection{GeoTrajGAN}

We did not test \gls{gtg} on \gls{mnist-seq} as we already knew from our experiments with \gls{gpg} (\refer \secref{sec_pointgen}), that the baseline model can capture the point distribution of the considered datasets (\refer \figref{fig_geopointgan}).
For both \gls{fs} and Geolife, we use a latent dimension of $256$ for both the generator and the discriminator. 
The main architecture differences to GeoPointGAN~\cite{GeoPointGAN} are the use of layer norm instead of batch norm in both models, an \gls{lstm} with hidden size $64$ in the generator, and the usage of \gls{stn} in the point-level discriminator.
Moreover, we add a sequential discriminator consisting of an \gls{lstm} with hidden size $64$ followed by two \gls{fc} layers.
The feedback of both discriminators is combined via addition.
A standard \gls{bce}-loss is used as the loss function. 
Both the generator and the combined discriminator are trained with the same frequency ($n\_critic$ = 1) using the \textit{\gls{adamw}} optimizer with a learning rate of 1E-5, $\beta_1 = 0.5$, $\beta_2=0.999$ and a batch size of $256$. 
We train the model on both datasets for $1000$ epochs as \gls{gtg} only produced a reasonable point distribution for \gls{fs} despite the larger size of the Geolife dataset (yielding more updates per epoch).
All other (hyper-)parameters are kept as their respective default values.

\end{document}